\documentclass{article}
\usepackage{amssymb,amsmath,amsfonts,graphicx,hyperref}
\newcommand{\mcm}[3]{\newcommand{#1}[#2]{{\ensuremath{#3}}}}

\usepackage{latexsym}
\usepackage{amssymb}

\mcm{\blank}{0}{(\emptybk)} \mcm{\dashbk}{0}{\mbox{---}}
\mcm{\emptybk}{0}{\:\:} \mcm{\hyph}{0}{\mbox{-}}

\mcm{\diagspace}{0}{\mbox{\hspace{2em}}}

\mcm{\cat}{1}{\mc{#1}} \mcm{\fcat}{1}{\mb{#1}}
\mcm{\mc}{1}{\mathcal{#1}} \mcm{\mr}{1}{\mathrm{#1}}
\mcm{\mi}{1}{\mathit{#1}} \mcm{\mb}{1}{\mathbf{#1}}
\mcm{\scat}{1}{\Bbb{#1}} \mcm{\twid}{1}{\widetilde{#1}}

\mcm{\elt}{0}{\in} \mcm{\sub}{0}{\,\subseteq\,}
\mcm{\such}{0}{\:|\:} \mcm{\without}{0}{\setminus}

\mcm{\atsr}{0}{\Box} \mcm{\eqv}{0}{\,\simeq\,}
\mcm{\iso}{0}{\,\cong\,}
\mcm{\of}{0}{\raisebox{0.2mm}{\ensuremath{\scriptstyle\circ}}}

\mcm{\bdry}{0}{\partial}

\mcm{\Bee}{0}{\cat{B}} \mcm{\Beep}{0}{\cat{B'}}
\mcm{\Eee}{0}{\cat{E}} \mcm{\Eeep}{0}{\cat{E'}}

\mcm{\Ess}{0}{\cat{S}} \mcm{\Tee}{0}{\cat{T}}
\mcm{\Teep}{0}{\cat{T'}} \mcm{\Stee}{0}{\scat{T}}
\mcm{\Steep}{0}{\scat{T'}}

\mcm{\blbk}{0}{\blank^{\blob}}
\mcm{\blob}{0}{\scriptscriptstyle{\bullet}}
\mcm{\stbk}{0}{\blank^{*}} \mcm{\ubl}{0}{{}^{\blob}}
\mcm{\ust}{0}{{}^{*}}

\mcm{\Cartpr}{0}{\pr{\Eee}{T}} \mcm{\Cartprp}{0}{\pr{\Eeep}{T'}}
\mcm{\Mnd}{0}{\triple{T}{\eta}{\mu}}
\mcm{\Zeropr}{0}{\pr{\Set}{\id}}

\mcm{\dopset}{0}{\ftrcat{\Delta^{\op}}{\Set}}
\mcm{\tropset}{0}{\ftrcat{\fcat{TR}^{\op}}{\Set}}

\mcm{\cod}{0}{\mr{cod}} \mcm{\dom}{0}{\mr{dom}}
\mcm{\End}{0}{\mr{End}} \mcm{\Hom}{0}{\mr{Hom}}
\mcm{\ob}{0}{\mr{ob}\,} \mcm{\op}{0}{\mr{op}}

\mcm{\comp}{0}{\mi{comp}} \mcm{\id}{0}{\mi{id}}
\mcm{\ids}{0}{\mi{ids}} \mcm{\mult}{0}{\mi{mult}}
\mcm{\unit}{0}{\mi{unit}}

\mcm{\Ab}{0}{\fcat{Ab}} \mcm{\Alg}{0}{\fcat{Alg}}
\mcm{\Bim}{1}{\fcat{Bim}(#1)} \mcm{\Cat}{0}{\fcat{Cat}}
\mcm{\Cay}{0}{\fcat{Cay}} \mcm{\Cpn}{1}{\pr{\Set/S_{#1}}{T_{#1}}}
\mcm{\fc}{0}{\fcat{fc}} \mcm{\fm}{0}{\fcat{fm}}
\mcm{\Graph}{0}{\fcat{Graph}} \mcm{\Gy}{0}{\fcat{Gy}}
\mcm{\Hpn}{1}{\pr{\Eee_{#1}}{P_{#1}}} \mcm{\Mon}{0}{\mb{Mon}}
\mcm{\Multicat}{0}{\fcat{Multicat}} \mcm{\One}{0}{\fcat{1}}
\mcm{\PD}{1}{\fcat{PD}_{#1}} \mcm{\Prof}{0}{\fcat{Prof}}
\mcm{\Set}{0}{\fcat{Set}} \mcm{\Span}{0}{\fcat{Span}}
\mcm{\Ssq}{0}{\fcat{Ssq}} \mcm{\Struc}{0}{\fcat{Struc}}
\mcm{\Sym}{0}{\fcat{Sym}} \mcm{\TR}{1}{\fcat{TR}(#1)}
\mcm{\Tr}{0}{\fcat{Tr}} \mcm{\Twocat}{0}{\fcat{2\hyph\Cat}}

\mcm{\integers}{0}{\mathbb{Z}}

\mcm{\range}{2}{#1,\,\ldots\,,#2}
\mcm{\bftuple}{2}{\tuplebts{\range{#1}{#2}}}
\mcm{\tuple}{3}{\tuplebts{\range{#1,#2}{#3}}}
\mcm{\rttuple}{1}{\tuplebts{\,\ldots\,,#1}}
\mcm{\abftuple}{2}{\atuplebts{\range{#1}{#2}}}
\mcm{\atuple}{3}{\atuplebts{\range{#1,#2}{#3}}}
\mcm{\arttuple}{1}{\atuplebts{\,\ldots\,,#1}}
\mcm{\sqbftuple}{2}{\obt\range{#1}{#2}\cbt}
\mcm{\pr}{2}{\tuplebts{#1,#2}}
\mcm{\triple}{3}{\tuplebts{#1,#2,#3}}

\mcm{\eend}{2}{#1[#2]} \mcm{\ehom}{3}{#1[#2,#3]}
\mcm{\ftrcat}{2}{[#1,#2]} \mcm{\homset}{3}{#1(#2,#3)}
\mcm{\multihom}{3}{#1(#2;#3)}
\mcm{\relhom}{5}{#1_{#2}(\range{#3}{#4};#5)}

\mcm{\go}{0}{\rTo} \mcm{\goby}{1}{\rTo^{#1}}
\mcm{\goesto}{0}{\,\longmapsto\,} \mcm{\goiso}{0}{\goby{\diso}}
\mcm{\monic}{0}{\rMonic} \mcm{\og}{0}{\lTo}
\mcm{\ogby}{1}{\lTo^{#1}}

\mcm{\gph}{2}{\spn{#1}{T #2}{#2}} \mcm{\graph}{4}{\spaan{#1}{T
#2}{#2}{#3}{#4}} \mcm{\oppair}{2}{\stackrel{\rTo^{#1}}{\lTo_{#2}}}
\mcm{\parpair}{2}{\stackrel{\rTo^{#1}}{\rTo_{#2}}}
\mcm{\spn}{3}{#2 \og #1 \go #3} \mcm{\spaan}{5}{#2 \ogby{#4} #1
\goby{#5} #3}

\mcm{\bktdvslob}{3}
    {\left(
    \begin{diagram}[height=1.5em]
    #1      \\
    \dTo>{\,#2} \\
    #3      \\
    \end{diagram}
    \right)}
\mcm{\slob}{3}{(#1 \goby{#2} #3)} \mcm{\vslob}{3}
    {\left.
    \begin{diagram}[height=1.5em]
    #1      \\
    \dTo>{\,#2} \\
    #3      \\
    \end{diagram}
    \right.}

\newenvironment{tree}
    {\begin{diagram}[height=1em,width=.75em,abut,noPS,tight]}
    {\end{diagram}}

\mcm{\enode}{0}{\circ}

\mcm{\nl}{1}{\stackrel{\textstyle #1}{\node}}
\mcm{\node}{0}{\bullet}

\mcm{\utree}{0}{\node}

\mcm{\diso}{0}{\sim}

\mcm{\vdiso}{0}{\wr}

\mcm{\nat}{0}{\mathbb{N}}

\mcm{\Onepr}{0}{\pr{\Graph}{\fc}}
\newlength{\nllwidth}
\newlength{\nllheight}
\newcommand{\stackbelow}[2]{%
\settowidth{\nllwidth}{\ensuremath{#1}\ensuremath{#2}}%
\settoheight{\nllheight}{\ensuremath{#2}}%
\addtolength{\nllheight}{.3ex}%
\mbox{%
\ensuremath{#1}%
\hspace{-.5\nllwidth}%
\raisebox{-1\nllheight}{\ensuremath{#2}}}}
\mcm{\nlal}{2}{\stackbelow{\nl{#1}}{#2}}
\mcm{\nll}{1}{\stackbelow{\node}{#1}} \mcm{\wun}{0}{\fcat{1}}
\mcm{\atuplebts}{1}{\langle #1 \rangle} \mcm{\tuplebts}{1}{(#1)}
\mcm{\bo}{0}{(} \mcm{\bc}{0}{)}
\mcm{\UBilax}{0}{\fcat{UBicat}_\mr{lax}}
\mcm{\UBiwk}{0}{\fcat{UBicat}_\mr{wk}}
\mcm{\UBistr}{0}{\fcat{UBicat}_\mr{str}}
\mcm{\Bilax}{0}{\fcat{Bicat}_\mr{lax}}
\mcm{\Biwk}{0}{\fcat{Bicat}_\mr{wk}}
\mcm{\Bistr}{0}{\fcat{Bicat}_\mr{str}} \mcm{\rotsub}{0}{\cup
\raisebox{0.1em}{$\scriptstyle{|}$}} \mcm{\pd}{0}{\fcat{pd}}
\mcm{\rep}{1}{\widehat{#1}} \mcm{\ovln}{1}{\overline{#1}}
\mcm{\Gph}{0}{\fcat{Gph}} \mcm{\tr}{0}{\fcat{tr}}

\mcm{\ladj}{0}{\,\dashv\,} \mcm{\zeropd}{0}{\node}
    {\end{diagram}}
\mcm{\END}{0}{\fcat{End}} \mcm{\HOM}{0}{\fcat{Hom}}

\newlength{\gwidth} 
\newlength{\gvert}  
\newlength{\gdrop}  
\newlength{\gbaredrop}  
\newlength{\goffset}    
\newlength{\gtemp}  

\newcommand{\present}[1]{%
\makebox[1\gwidth]{%
\rule[-1\gdrop]{0ex}{1\gvert}%
\raisebox{-1\gbaredrop}{#1}}}

\newcommand{\presentl}[1]{%
\makebox[1\gwidth][l]{%
\rule[-1\gdrop]{0ex}{1\gvert}%
\raisebox{-1\gbaredrop}{#1}}}

\newcommand{\presentr}[1]{%
\makebox[1\gwidth][r]{%
\rule[-1\gdrop]{0ex}{1\gvert}%
\raisebox{-1\gbaredrop}{#1}}}

\newcommand{\ginitdims}[2]{
\setlength{\unitlength}{1em}
\setlength{\goffset}{.25\unitlength}
\setlength{\gwidth}{#1\unitlength}
\setlength{\gvert}{#2\unitlength}
\setlength{\gdrop}{.5\gvert}
\addtolength{\gdrop}{-1\goffset}
\setlength{\gbaredrop}{1\gdrop}
\addtolength{\gvert}{.6\unitlength}
\addtolength{\gdrop}{.3\unitlength}}    

\newcommand{\cinitdims}[2]{
\setlength{\unitlength}{1em}
\setlength{\goffset}{.35\unitlength}
\setlength{\gwidth}{#1\unitlength}
\setlength{\gvert}{#2\unitlength}
\setlength{\gdrop}{.5\gvert}
\addtolength{\gdrop}{-1\goffset}
\setlength{\gbaredrop}{1\gdrop}
\addtolength{\gvert}{.6\unitlength}
\addtolength{\gdrop}{.3\unitlength}}    

\newcommand{\gsinitdims}[2]{
\setlength{\unitlength}{0.5em}
\setlength{\goffset}{.25\unitlength}
\setlength{\gwidth}{#1\unitlength}
\setlength{\gvert}{#2\unitlength}
\setlength{\gdrop}{.5\gvert}
\addtolength{\gdrop}{-1\goffset}
\setlength{\gbaredrop}{1\gdrop}
\addtolength{\gvert}{.6\unitlength}
\addtolength{\gdrop}{.3\unitlength}}    

\newcommand{\sidespic}[1]{%
\settowidth{\gtemp}{\ensuremath{#1}}%
\addtolength{\gwidth}{1\gtemp}}

\newcommand{\abovepic}[1]{%
\settoheight{\gtemp}{\ensuremath{#1}}%
\addtolength{\gvert}{1\gtemp}%
\settodepth{\gtemp}{\ensuremath{#1}}%
\addtolength{\gvert}{1\gtemp}}

\newcommand{\belowpic}[1]{%
\settoheight{\gtemp}{\ensuremath{#1}}%
\addtolength{\gvert}{1\gtemp}%
\addtolength{\gdrop}{1\gtemp}%
\settodepth{\gtemp}{\ensuremath{#1}}%
\addtolength{\gvert}{1\gtemp}%
\addtolength{\gdrop}{1\gtemp}}

\newcommand{\cell}[4]{\put(#1,#2){\makebox(0,0)[#3]{\ensuremath{#4}}}}
\mcm{\zmark}{0}{\scriptstyle{\bullet}}

\newcommand{\pregfst}[1]{%
\begin{picture}(0.5,0.2)(-0.5,-0.2)%
\cell{-0.1}{-0.2}{tr}{#1}%
\cell{0}{0}{c}{\zmark}%
\end{picture}}

\mcm{\gfst}{1}{%
\ginitdims{0.5}{0.4}%
\sidespic{#1}%
\belowpic{#1}%
\presentr{\pregfst{#1}}}

\newcommand{\preglst}[1]{%
\begin{picture}(0.5,0.2)(0,-0.2)%
\cell{0.1}{-0.2}{tl}{#1}%
\cell{0.05}{0}{c}{\zmark}%
\end{picture}}

\mcm{\glst}{1}{%
\ginitdims{.5}{.4}%
\sidespic{#1}%
\belowpic{#1}%
\presentl{\preglst{#1}}}

\newcommand{\preglft}[1]{%
\begin{picture}(0,0.2)(0,-0.2)%
\cell{-0.1}{-0.2}{tr}{#1}%
\cell{0.05}{0}{c}{\zmark}%
\end{picture}}

\mcm{\glft}{1}{%
\ginitdims{0}{.4}%
\belowpic{#1}%
\present{\preglft{#1}}}

\newcommand{\pregrgt}[1]{%
\begin{picture}(0,0.2)(0,-0.2)%
\cell{0.1}{-0.2}{tl}{#1}%
\cell{0.05}{0}{c}{\zmark}%
\end{picture}}

\mcm{\grgt}{1}{%
\ginitdims{0}{.4}%
\belowpic{#1}%
\present{\pregrgt{#1}}}

\newcommand{\pregblw}[1]{%
\begin{picture}(0,0.3)(0,-0.3)
\cell{0}{-0.3}{t}{#1}%
\cell{0.05}{0}{c}{\zmark}%
\end{picture}}

\mcm{\gblw}{1}{%
\ginitdims{0}{.6}%
\belowpic{#1}%
\present{\pregblw{#1}}}

\newcommand{\pregfbw}[1]{%
\begin{picture}(0,0.65)(0,-0.65)
\cell{0}{-0.65}{t}{#1}%
\cell{0.05}{0}{c}{\zmark}%
\end{picture}}

\mcm{\gfbw}{1}{%
\ginitdims{0}{1.3}%
\belowpic{#1}%
\present{\pregfbw{#1}}}

\newcommand{\pregzero}[1]{%
\begin{picture}(0.8,0.4)(-0.4,-0.4)
\cell{0}{-0.4}{t}{#1}%
\cell{0}{0}{c}{\zmark}%
\end{picture}}

\mcm{\gzero}{1}{%
\ginitdims{0.8}{.6}%
\belowpic{#1}%
\sidespic{#1}%
\present{\pregzero{#1}}}

\newcommand{\pregone}[1]{%
\begin{picture}(5,0.4)(0,-0.2)%
\cell{2.5}{0.2}{b}{#1}%
\put(0,0){\vector(1,0){5}}%
\end{picture}}

\mcm{\gone}{1}{%
\ginitdims{5}{0.4}%
\abovepic{#1}%
\present{\pregone{#1}}}

\newcommand{\pregtwo}[3]{%
\begin{picture}(5,3.4)(0,-0.2)%
\cell{2.5}{3.2}{b}{#1}%
\cell{2.5}{-.2}{t}{#2}%
\cell{2.7}{1.5}{l}{#3}%
\qbezier(0,1.5)(2.5,4.5)(5,1.5)%
\qbezier(0,1.5)(2.5,-1.5)(5,1.5)%
\put(5,1.5){\vector(1,-1){0}}%
\put(5,1.5){\vector(1,1){0}}%
\put(2.5,2.5){\vector(0,-1){2}}%
\end{picture}}

\mcm{\gtwo}{3}{%
\ginitdims{5}{3.4}%
\abovepic{#1}%
\belowpic{#2}%
\present{\pregtwo{#1}{#2}{#3}}}

\newcommand{\pregthree}[5]{%
\begin{picture}(5,5.4)(0,-1.2)%
\cell{2.5}{4.2}{b}{#1}%
\cell{1.5}{1.7}{b}{#2}%
\cell{2.5}{-1.2}{t}{#3}%
\cell{2.7}{2.75}{l}{#4}%
\cell{2.7}{0.25}{l}{#5}%
\qbezier(0,1.5)(2.5,6.5)(5,1.5)%
\qbezier(0,1.5)(2.5,-3.5)(5,1.5)%
\put(0,1.5){\vector(1,0){5}}%
\put(2.5,3.5){\vector(0,-1){1.5}}%
\put(2.5,1){\vector(0,-1){1.5}}%
\put(5,1.5){\vector(1,-3){0}}%
\put(5,1.5){\vector(1,3){0}}%
\end{picture}}

\mcm{\gthree}{5}{%
\ginitdims{5}{5.4}%
\abovepic{#1}%
\belowpic{#3}%
\present{\pregthree{#1}{#2}{#3}{#4}{#5}}}

\newcommand{\pregfour}[7]{%
\begin{picture}(5,8.4)(0,-2.7)%
\cell{2.5}{5.7}{b}{#1}%
\cell{1.5}{2.8}{b}{#2}%
\cell{1.5}{0.2}{t}{#3}%
\cell{2.5}{-2.7}{t}{#4}%
\cell{2.7}{4.25}{l}{#5}%
\cell{2.7}{1.5}{l}{#6}%
\cell{2.7}{-1.25}{l}{#7}%
\qbezier(0,1.5)(2.5,9.5)(5,1.5)%
\qbezier(0,1.5)(2.5,4)(5,1.5)%
\qbezier(0,1.5)(2.5,-1)(5,1.5)%
\qbezier(0,1.5)(2.5,-6.5)(5,1.5)%
\put(2.5,5.25){\vector(0,-1){2}}%
\put(2.5,2.5){\vector(0,-1){2}}%
\put(2.5,-0.25){\vector(0,-1){2}}%
\put(5,1.5){\vector(1,-4){0}}%
\put(5,1.5){\vector(4,-3){0}}%
\put(5,1.5){\vector(4,3){0}}%
\put(5,1.5){\vector(1,4){0}}%
\end{picture}}

\mcm{\gfour}{7}{%
\ginitdims{5}{8.4}%
\abovepic{#1}%
\belowpic{#4}%
\present{\pregfour{#1}{#2}{#3}{#4}{#5}{#6}{#7}}}

\newcommand{\pregthreecell}[5]{%
\begin{picture}(8,5)(-4,-2.5)%
\cell{0}{2.5}{b}{#1}%
\cell{0}{-2.5}{t}{#2}%
\cell{-1.7}{0}{r}{#3}%
\cell{1.7}{0}{l}{#4}%
\cell{0}{0.2}{b}{#5}%
\qbezier(-4,0)(0,4.2)(4,0)%
\qbezier(-4,0)(0,-4.2)(4,0)%
\qbezier(-0.5,1.8)(-2.5,0)(-0.5,-1.8)%
\qbezier(0.5,1.8)(2.5,0)(0.5,-1.8)%
\put(-1,0){\vector(1,0){2}}%
\put(4,0){\vector(1,-1){0}}%
\put(4,0){\vector(1,1){0}}%
\put(-0.5,-1.8){\vector(1,-1){0}}%
\put(0.5,-1.8){\vector(-1,-1){0}}%
\end{picture}}

\mcm{\gthreecell}{5}{%
\ginitdims{8}{5}%
\abovepic{#1}%
\belowpic{#2}%
\present{\pregthreecell{#1}{#2}{#3}{#4}{#5}}}

\newcommand{\pregthreecellu}{%
\begin{picture}(5,3.4)(-0.5,-0.2)%
\qbezier(-.5,1.5)(2,4.5)(4.5,1.5)%
\qbezier(-.5,1.5)(2,-1.5)(4.5,1.5)%
\qbezier(1.5,2.7)(0.5,1.5)(1.5,0.3)%
\qbezier(2.5,2.7)(3.5,1.5)(2.5,0.3)%
\put(1.3,1.5){\vector(1,0){1.4}}%
\put(4.5,1.5){\vector(1,-1){0}}%
\put(4.5,1.5){\vector(1,1){0}}%
\put(1.5,0.3){\vector(2,-3){0}}%
\put(2.5,0.3){\vector(-2,-3){0}}%
\end{picture}}

\mcm{\gthreecellu}{0}{%
\ginitdims{5}{3.4}%
\present{\pregthreecellu}}

\newcommand{\pregtwocentre}[3]{%
\begin{picture}(5,3.4)(0,-0.2)%
\cell{2.5}{3.2}{b}{#1}%
\cell{2.5}{-.2}{t}{#2}%
\cell{2.5}{1.5}{c}{#3}%
\qbezier(0,1.5)(2.5,4.5)(5,1.5)%
\qbezier(0,1.5)(2.5,-1.5)(5,1.5)%
\put(5,1.5){\vector(1,-1){0}}%
\put(5,1.5){\vector(1,1){0}}%
\put(2.5,2.5){\vector(0,-1){2}}%
\end{picture}}

\mcm{\gtwocentre}{3}{%
\ginitdims{5}{3.4}%
\abovepic{#1}%
\belowpic{#2}%
\present{\pregtwocentre{#1}{#2}{#3}}}

\newcommand{\pregspecialone}[9]{%
\begin{picture}(8,8)(-4,-4)%
\cell{0}{3.9}{b}{#1}%
\cell{-2}{-0.2}{t}{#2}%
\cell{0}{-3.9}{t}{#3}%
\cell{-1.5}{1.1}{r}{#4}%
\cell{0.2}{1.5}{l}{#5}%
\cell{1.5}{1.1}{l}{#6}%
\cell{0.2}{-2}{l}{#7}%
\cell{-0.9}{2.3}{b}{#8}%
\cell{0.9}{2.3}{b}{#9}%
\qbezier(-4,0)(0,8)(4,0)%
\qbezier(-4,0)(0,-8)(4,0)%
\qbezier(-0.5,3.4)(-3.5,2)(-0.5,0.6)%
\qbezier(0.5,3.4)(3.5,2)(0.5,0.6)%
\put(-4,0){\vector(1,0){8}}%
\put(0,3.4){\vector(0,-1){2.8}}%
\put(0,-0.8){\vector(0,-1){2.4}}%
\put(-1.5,2.2){\vector(1,0){1.2}}%
\put(0.3,2.2){\vector(1,0){1.2}}%
\put(4,0){\vector(1,-2){0}}%
\put(4,0){\vector(1,2){0}}%
\put(-0.5,0.6){\vector(2,-1){0}}%
\put(0.5,0.6){\vector(-2,-1){0}}%
\end{picture}}

\mcm{\gspecialone}{9}{%
\ginitdims{8}{8}%
\abovepic{#1}%
\belowpic{#3}%
\present{\pregspecialone{#1}{#2}{#3}{#4}{#5}{#6}{#7}{#8}{#9}}}

\newcommand{\pregspecialtwo}{%
\begin{picture}(5,3.4)(0,-0.2)%
\qbezier(0,1.5)(2.5,4.5)(5,1.5)%
\qbezier(0,1.5)(2.5,-1.5)(5,1.5)%
\qbezier(1.7,2.5)(0,1.5)(1.7,0.5)%
\qbezier(3.3,2.5)(5,1.5)(3.3,0.5)%
\put(5,1.5){\vector(1,-1){0}}%
\put(5,1.5){\vector(1,1){0}}%
\put(1.7,0.5){\vector(3,-2){0}}%
\put(3.3,0.5){\vector(-3,-2){0}}%
\put(2.5,2.5){\vector(0,-1){2}}%
\put(1.2,1.5){\vector(1,0){1}}%
\put(2.8,1.5){\vector(1,0){1}}%
\end{picture}}

\mcm{\gspecialtwo}{0}{%
\ginitdims{5}{3.4}%
\present{\pregspecialtwo}}

\newcommand{\pregspecialthree}{%
\begin{picture}(5,5.4)(0,-1.2)%
\qbezier(0,1.5)(2.5,6.5)(5,1.5)%
\qbezier(0,1.5)(2.5,-3.5)(5,1.5)%
\qbezier(2,3.5)(1,2.75)(2,2)%
\qbezier(3,3.5)(4,2.75)(3,2)%
\qbezier(2,1)(1,0.25)(2,-0.5)%
\qbezier(3,1)(4,0.25)(3,-0.5)%
\put(0,1.5){\vector(1,0){5}}%
\put(1.5,2.75){\vector(1,0){2}}%
\put(1.5,0.25){\vector(1,0){2}}%
\put(5,1.5){\vector(1,-3){0}}%
\put(5,1.5){\vector(1,3){0}}%
\put(2,2){\vector(1,-1){0}}%
\put(3,2){\vector(-1,-1){0}}%
\put(2,-0.5){\vector(1,-1){0}}%
\put(3,-0.5){\vector(-1,-1){0}}%
\end{picture}}

\mcm{\gspecialthree}{0}{%
\ginitdims{5}{5.4}%
\present{\pregspecialthree}}

\newcommand{\pregonew}[1]{%
\begin{picture}(8,0.4)(0,-0.2)%
\cell{4}{0.2}{b}{#1}%
\put(0,0){\vector(1,0){8}}%
\end{picture}}

\mcm{\gonew}{1}{%
\ginitdims{8}{0.4}%
\abovepic{#1}%
\present{\pregonew{#1}}}

\mcm{\gzersu}{0}{%
\gsinitdims{0}{.6}%
\present{\pregblw{}}}

\mcm{\gonesu}{0}{%
\gsinitdims{5}{0.4}%
\present{\pregone{}}}

\mcm{\gtwosu}{0}{%
\gsinitdims{5}{3.4}%
\present{\pregtwo{}{}{}}}

\mcm{\gthreesu}{0}{%
\gsinitdims{5}{5.4}%
\present{\pregthree{}{}{}{}{}}}

\mcm{\gfoursu}{0}{%
\gsinitdims{5}{8.4}%
\present{\pregfour{}{}{}{}{}{}{}}}

\newcommand{\precone}[1]{%
\begin{picture}(4.2,0.4)(-0.3,-0.2)%
\cell{1.8}{0.2}{b}{#1}%
\put(0,0){\vector(1,0){3.6}}%
\end{picture}}

\mcm{\cone}{1}{%
\cinitdims{4.2}{0.4}%
\abovepic{#1}%
\present{\precone{#1}}}

\mcm{\gfstsu}{0}{%
\gsinitdims{0.5}{0.4}%
\presentr{\pregfst{}}}

\mcm{\glstsu}{0}{%
\gsinitdims{0.5}{0.4}%
\presentl{\preglst{}}}

\newcommand{\prectwodbl}[3]%
{\begin{picture}(4.2,3.4)(-0.1,-0.2)%
\cell{2}{3.2}{b}{#1}%
\cell{2}{-0.2}{t}{#2}%
\cell{2.3}{1.5}{l}{#3}%
\qbezier(0,2)(2,4)(4,2)%
\qbezier(0,1)(2,-1)(4,1)%
\put(4,2){\vector(1,-1){0}}%
\put(4,1){\vector(1,1){0}}%
\put(1.9,2.5){\line(0,-1){1.8}}%
\put(2.1,2.5){\line(0,-1){1.8}}%
\cell{2.01}{0.4}{b}{\vee}%
\end{picture}}

\mcm{\ctwodbl}{3}{%
\cinitdims{4.2}{3.4}%
\abovepic{#1}%
\belowpic{#2}%
\present{\prectwodbl{#1}{#2}{#3}}}

\newcommand{\precthreedbl}[5]{%
\begin{picture}(4.2,5.4)(-0.1,-0.2)%
\cell{2}{5.2}{b}{#1}%
\cell{1}{2.7}{b}{#2}%
\cell{2}{-.2}{t}{#3}%
\cell{2.3}{3.75}{l}{#4}%
\cell{2.3}{1.25}{l}{#5}%
\qbezier(0,3)(2,7)(4,3)%
\qbezier(0,2)(2,-2)(4,2)%
\put(0,2.5){\vector(1,0){4}}%
\put(1.9,4.5){\line(0,-1){1.3}}%
\put(2.1,4.5){\line(0,-1){1.3}}%
\cell{2.01}{2.9}{b}{\vee}%
\put(1.9,2){\line(0,-1){1.3}}%
\put(2.1,2){\line(0,-1){1.3}}%
\cell{2.01}{0.4}{b}{\vee}%
\put(4,3){\vector(1,-3){0}}%
\put(4,2){\vector(1,3){0}}%
\end{picture}}

\mcm{\cthreedbl}{5}{%
\cinitdims{4.2}{5.4}%
\abovepic{#1}%
\belowpic{#3}%
\present{\precthreedbl{#1}{#2}{#3}{#4}{#5}}}

\newcommand{\precthreecelltrp}[5]{%
\begin{picture}(8.2,5)(-4.1,-2.5)%
\cell{0}{2.5}{b}{#1}%
\cell{0}{-2.5}{t}{#2}%
\cell{-1.8}{0}{r}{#3}%
\cell{1.8}{0}{l}{#4}%
\cell{0}{0.3}{b}{#5}%
\qbezier(-4,0.5)(0,4)(4,0.5)%
\qbezier(-4,-0.5)(0,-4)(4,-0.5)%
\qbezier(-0.6,2)(-2.6,0)(-0.6,-2)%
\qbezier(-0.4,2)(-2.4,0)(-0.5,-1.9)%
\cell{-0.6}{-2}{b}{\lrcorner}%
\qbezier(0.4,2)(2.4,0)(0.5,-1.9)%
\qbezier(0.6,2)(2.6,0)(0.6,-2)%
\cell{0.65}{-2}{b}{\llcorner}%
\put(-1,0.15){\line(1,0){1.7}}%
\put(-1,0){\line(1,0){2}}%
\put(-1,-0.15){\line(1,0){1.7}}%
\cell{1.15}{0}{r}{>}%
\put(4,0.5){\vector(1,-1){0}}%
\put(4,-0.5){\vector(1,1){0}}%
\end{picture}}

\mcm{\cthreecelltrp}{5}{%
\cinitdims{8.2}{5}%
\abovepic{#1}%
\belowpic{#2}%
\present{\precthreecelltrp{#1}{#2}{#3}{#4}{#5}}}

\newcommand{\prectwo}[3]%
{\begin{picture}(4.2,3.4)(-0.1,-0.2)%
\cell{2}{3.2}{b}{#1}%
\cell{2}{-0.2}{t}{#2}%
\cell{2.2}{1.5}{l}{#3}%
\qbezier(0,2)(2,4)(4,2)%
\qbezier(0,1)(2,-1)(4,1)%
\put(4,2){\vector(1,-1){0}}%
\put(4,1){\vector(1,1){0}}%
\put(2,2.5){\vector(0,-1){2}}%
\end{picture}}

\mcm{\ctwo}{3}{%
\cinitdims{4.2}{3.4}%
\abovepic{#1}%
\belowpic{#2}%
\present{\prectwo{#1}{#2}{#3}}}

\newcommand{\precthree}[5]{%
\begin{picture}(4.2,5.4)(-0.1,-0.2)%
\cell{2}{5.2}{b}{#1}%
\cell{1}{2.7}{b}{#2}%
\cell{2}{-.2}{t}{#3}%
\cell{2.2}{3.75}{l}{#4}%
\cell{2.2}{1.25}{l}{#5}%
\qbezier(0,3)(2,7)(4,3)%
\qbezier(0,2)(2,-2)(4,2)%
\put(0,2.5){\vector(1,0){4}}%
\put(2,4.5){\vector(0,-1){1.5}}%
\put(2,2){\vector(0,-1){1.5}}%
\put(4,3){\vector(1,-3){0}}%
\put(4,2){\vector(1,3){0}}%
\end{picture}}

\mcm{\cthree}{5}{%
\cinitdims{4.2}{5.4}%
\abovepic{#1}%
\belowpic{#3}%
\present{\precthree{#1}{#2}{#3}{#4}{#5}}}

\newcommand{\prectwoop}[3]%
{\begin{picture}(4.2,3.4)(-0.1,-0.2)%
\cell{2}{3.2}{b}{#1}%
\cell{2}{-0.2}{t}{#2}%
\cell{2.2}{1.5}{l}{#3}%
\qbezier(0,2)(2,4)(4,2)%
\qbezier(0,1)(2,-1)(4,1)%
\put(0,2){\vector(-1,-1){0}}%
\put(0,1){\vector(-1,1){0}}%
\put(2,2.5){\vector(0,-1){2}}%
\end{picture}}

\mcm{\ctwoop}{3}{%
\cinitdims{4.2}{3.4}%
\abovepic{#1}%
\belowpic{#2}%
\present{\prectwoop{#1}{#2}{#3}}}

\newcommand{\prectwopar}[4]{%
\begin{picture}(4.2,3.4)(-0.1,-0.2)%
\cell{2}{3.2}{b}{#1}%
\cell{2}{-0.2}{t}{#2}%
\cell{1.6}{1.5}{r}{#3}%
\cell{2.4}{1.5}{l}{#4}%
\qbezier(0,2)(2,4)(4,2)%
\qbezier(0,1)(2,-1)(4,1)%
\put(4,2){\vector(1,-1){0}}%
\put(4,1){\vector(1,1){0}}%
\put(1.8,2.5){\vector(0,-1){2}}%
\put(2.2,2.5){\vector(0,-1){2}}%
\end{picture}}

\mcm{\ctwopar}{4}{%
\cinitdims{4.2}{3.4}%
\abovepic{#1}%
\belowpic{#2}%
\present{\prectwopar{#1}{#2}{#3}{#4}}}

\newcommand{\precthreein}[5]{%
\begin{picture}(4.2,5.4)(-0.1,-0.2)%
\cell{2}{5.2}{b}{#1}%
\cell{1}{2.7}{b}{#2}%
\cell{2}{-.2}{t}{#3}%
\cell{2.2}{3.75}{l}{#4}%
\cell{2.2}{1.25}{l}{#5}%
\qbezier(0,3)(2,7)(4,3)%
\qbezier(0,2)(2,-2)(4,2)%
\put(0,2.5){\vector(1,0){4}}%
\put(2,4.5){\vector(0,-1){1.5}}%
\put(2,0.5){\vector(0,1){1.5}}%
\put(4,3){\vector(1,-3){0}}%
\put(4,2){\vector(1,3){0}}%
\end{picture}}

\mcm{\cthreein}{5}{%
\cinitdims{4.2}{5.4}%
\abovepic{#1}%
\belowpic{#3}%
\present{\precthreein{#1}{#2}{#3}{#4}{#5}}}

\newcommand{\precthreecell}[5]{%
\begin{picture}(8.2,5)(-4.1,-2.5)%
\cell{0}{2.5}{b}{#1}%
\cell{0}{-2.5}{t}{#2}%
\cell{-1.7}{0}{r}{#3}%
\cell{1.7}{0}{l}{#4}%
\cell{0}{0.2}{b}{#5}%
\qbezier(-4,0.5)(0,4)(4,0.5)%
\qbezier(-4,-0.5)(0,-4)(4,-0.5)%
\qbezier(-0.5,2)(-2.5,0)(-0.5,-2)%
\qbezier(0.5,2)(2.5,0)(0.5,-2)%
\put(-1,0){\vector(1,0){2}}%
\put(4,0.5){\vector(1,-1){0}}%
\put(4,-0.5){\vector(1,1){0}}%
\put(-0.5,-2){\vector(1,-1){0}}%
\put(0.5,-2){\vector(-1,-1){0}}%
\end{picture}}

\mcm{\cthreecell}{5}{%
\cinitdims{8.2}{5}%
\abovepic{#1}%
\belowpic{#2}%
\present{\precthreecell{#1}{#2}{#3}{#4}{#5}}}

\newcommand{\precthreecellpar}[6]{%
\begin{picture}(8.2,5)(-4.1,-2.5)%
\cell{0}{2.5}{b}{#1}%
\cell{0}{-2.5}{t}{#2}%
\cell{-1.7}{0}{r}{#3}%
\cell{1.7}{0}{l}{#4}%
\cell{0}{0.4}{b}{#5}%
\cell{0}{-0.4}{t}{#6}%
\qbezier(-4,0.5)(0,4)(4,0.5)%
\qbezier(-4,-0.5)(0,-4)(4,-0.5)%
\qbezier(-0.5,2)(-2.5,0)(-0.5,-2)%
\qbezier(0.5,2)(2.5,0)(0.5,-2)%
\put(-1,0.2){\vector(1,0){2}}%
\put(-1,-0.2){\vector(1,0){2}}%
\put(4,0.5){\vector(1,-1){0}}%
\put(4,-0.5){\vector(1,1){0}}%
\put(-0.5,-2){\vector(1,-1){0}}%
\put(0.5,-2){\vector(-1,-1){0}}%
\end{picture}}

\mcm{\cthreecellpar}{6}{%
\cinitdims{8.2}{5}%
\abovepic{#1}%
\belowpic{#2}%
\present{\precthreecellpar{#1}{#2}{#3}{#4}{#5}{#6}}}

\newcommand{\prectwov}[5]{%
\begin{picture}(3.4,4.2)(0.8,0.9)%
\cell{2.5}{5.1}{b}{#1}%
\cell{2.5}{0.9}{t}{#2}%
\cell{0.8}{3}{r}{#3}%
\cell{4.2}{3}{l}{#4}%
\cell{2.5}{3.2}{b}{#5}%
\qbezier(2,5)(0,3)(2,1)%
\qbezier(3,5)(5,3)(3,1)%
\put(2,1){\vector(1,-1){0}}%
\put(3,1){\vector(-1,-1){0}}%
\put(1.5,3){\vector(1,0){2}}%
\end{picture}}

\mcm{\ctwov}{5}{%
\cinitdims{3.4}{4.2}%
\abovepic{#1}%
\belowpic{#2}%
\sidespic{#3}%
\sidespic{#4}%
\present{\prectwov{#1}{#2}{#3}{#4}{#5}}}

\newcommand{\precthreecellv}[7]{%
\begin{picture}(5,8.2)(0.5,-1.6)%
\cell{3}{6.6}{b}{#1}%
\cell{3}{-1.6}{t}{#2}%
\cell{0.5}{2.5}{r}{#3}%
\cell{5.5}{2.5}{l}{#4}%
\cell{3}{4.2}{b}{#5}%
\cell{3}{0.8}{t}{#6}%
\cell{3.2}{2.5}{l}{#7}%
\qbezier(3.5,6.5)(7,2.5)(3.5,-1.5)%
\qbezier(2.5,6.5)(-1,2.5)(2.5,-1.5)%
\put(2.5,-1.5){\vector(1,-1){0}}%
\put(3.5,-1.5){\vector(-1,-1){0}}%
\qbezier(1,3)(3,5)(5,3)%
\qbezier(1,2)(3,0)(5,2)%
\put(5,3){\vector(1,-1){0}}%
\put(5,2){\vector(1,1){0}}%
\put(3,3.5){\vector(0,-1){2}}%
\end{picture}}

\mcm{\cthreecellv}{7}{%
\cinitdims{5}{8.2}%
\abovepic{#1}%
\belowpic{#2}%
\sidespic{#3}%
\sidespic{#4}%
\present{\precthreecellv{#1}{#2}{#3}{#4}{#5}{#6}{#7}}}

\newcommand{\pretopez}[2]{%
\begin{picture}(2.6,2.3)(-1.3,-2.2)%
\cell{0}{-2.2}{t}{#1}%
\cell{0}{-1.2}{c}{#2}%
\qbezier(0,0)(-2,-2)(0,-2)%
\qbezier(0,0)(2,-2)(0,-2)%
\put(0,0){\vector(-1,1){0}}%
\end{picture}}

\mcm{\topez}{2}{%
\ginitdims{2.6}{2.3}%
\belowpic{#1}%
\present{\pretopez{#1}{#2}}}

\newcommand{\pretopea}[3]{%
\begin{picture}(4,1.9)(-2,-0,2)%
\cell{0}{1.7}{b}{#1}%
\cell{0}{-0.2}{t}{#2}%
\cell{0}{0.7}{c}{#3}%
\qbezier(-2,0)(0,3)(2,0)%
\put(-2,0){\vector(1,0){4}}%
\put(2,0){\vector(2,-3){0}}%
\end{picture}}

\mcm{\topea}{3}{%
\ginitdims{4}{1.9}%
\abovepic{#1}%
\belowpic{#2}%
\present{\pretopea{#1}{#2}{#3}}}

\newcommand{\pretopeb}[4]{%
\begin{picture}(4,2.2)(-2,-0.2)%
\cell{-1.1}{1}{br}{#1}%
\cell{1.1}{1}{bl}{#2}%
\cell{0}{-0.2}{t}{#3}%
\cell{0}{0.8}{c}{#4}%
\put(-2,0){\vector(1,1){2}}%
\put(0,2){\vector(1,-1){2}}%
\put(-2,0){\vector(1,0){4}}%
\end{picture}}

\mcm{\topeb}{4}{%
\ginitdims{4}{2.2}%
\belowpic{#3}%
\present{\pretopeb{#1}{#2}{#3}{#4}}}

\newcommand{\pretopec}[5]{%
\begin{picture}(4,2.2)(-2,-0.2)%
\cell{-1.8}{1}{br}{#1}%
\cell{0}{2.2}{b}{#2}%
\cell{1.8}{1}{bl}{#3}%
\cell{0}{-0.2}{t}{#4}%
\cell{0}{0.8}{c}{#5}%
\put(-2,0){\vector(1,2){1}}%
\put(-1,2){\vector(1,0){2}}%
\put(1,2){\vector(1,-2){1}}%
\put(-2,0){\vector(1,0){4}}%
\end{picture}}

\mcm{\topec}{5}{%
\ginitdims{4}{2.2}%
\sidespic{#1}%
\abovepic{#2}%
\sidespic{#3}%
\belowpic{#4}%
\present{\pretopec{#1}{#2}{#3}{#4}{#5}}}

\newcommand{\pretoped}[6]{%
\begin{picture}(4,2.5)(-2,-0.2)%
\cell{-2}{0.6}{br}{#1}%
\cell{-0.7}{2.2}{br}{#2}%
\cell{0.7}{2.2}{bl}{#3}%
\cell{2}{0.6}{bl}{#4}%
\cell{0}{-0.2}{t}{#5}%
\cell{0}{0.8}{c}{#6}%
\put(-2,0){\vector(1,3){0.5}}%
\put(-1.5,1.5){\vector(3,2){1.5}}%
\put(0,2.5){\vector(3,-2){1.5}}%
\put(1.5,1.5){\vector(1,-3){0.5}}%
\put(-2,0){\vector(1,0){4}}%
\end{picture}}

\mcm{\toped}{6}{%
\ginitdims{4}{2.5}%
\sidespic{#1}%
\abovepic{#2}%
\abovepic{#3}%
\sidespic{#4}%
\belowpic{#5}%
\present{\pretoped{#1}{#2}{#3}{#4}{#5}{#6}}}

\newcommand{\pretopeq}[5]{%
\begin{picture}(4,2.5)(-2,-0.2)%
\cell{-2}{0.6}{br}{#1}%
\cell{-1}{2.2}{br}{#2}%
\cell{2}{0.6}{bl}{#3}%
\cell{0}{-0.2}{t}{#4}%
\cell{0}{0.8}{c}{#5}%
\put(-2,0){\vector(1,3){0.5}}%
\put(-1.5,1.5){\vector(1,1){1}}%
\cell{0.9}{2.3}{c}{\ddots}
\put(1.5,1.5){\vector(1,-3){0.5}}%
\put(-2,0){\vector(1,0){4}}%
\end{picture}}

\mcm{\topeq}{5}{%
\ginitdims{4}{2.5}%
\sidespic{#1}%
\abovepic{#2}%
\sidespic{#3}%
\belowpic{#4}%
\present{\pretopeq{#1}{#2}{#3}{#4}{#5}}}

\newcommand{\pretopebase}[1]{%
\begin{picture}(4,0.4)(0,-0.2)%
\cell{2}{0.2}{b}{#1}%
\put(0,0){\vector(1,0){4}}%
\end{picture}}

\mcm{\topebase}{1}{%
\ginitdims{4}{0.4}%
\abovepic{#1}%
\present{\pretopebase{#1}}}

\newcommand{\pretopezs}[2]{%
\begin{picture}(2.6,2.3)(-1.3,-2.2)%
\cell{0}{-2.2}{t}{#1}%
\cell{0}{-1.2}{c}{#2}%
\qbezier(0,0)(-2,-2)(0,-2)%
\qbezier(0,0)(2,-2)(0,-2)%
\end{picture}}

\mcm{\topezs}{2}{%
\ginitdims{2.6}{2.3}%
\belowpic{#1}%
\present{\pretopezs{#1}{#2}}}

\newcommand{\pretopeas}[3]{%
\begin{picture}(4,1.9)(-2,-0,2)%
\cell{0}{1.7}{b}{#1}%
\cell{0}{-0.2}{t}{#2}%
\cell{0}{0.7}{c}{#3}%
\qbezier(-2,0)(0,3)(2,0)%
\put(-2,0){\line(1,0){4}}%
\end{picture}}

\mcm{\topeas}{3}{%
\ginitdims{4}{1.9}%
\abovepic{#1}%
\belowpic{#2}%
\present{\pretopeas{#1}{#2}{#3}}}

\newcommand{\pretopebs}[4]{%
\begin{picture}(4,2.2)(-2,-0.2)%
\cell{-1.1}{1}{br}{#1}%
\cell{1.1}{1}{bl}{#2}%
\cell{0}{-0.2}{t}{#3}%
\cell{0}{0.8}{c}{#4}%
\put(-2,0){\line(1,1){2}}%
\put(0,2){\line(1,-1){2}}%
\put(-2,0){\line(1,0){4}}%
\end{picture}}

\mcm{\topebs}{4}{%
\ginitdims{4}{2.2}%
\belowpic{#3}%
\present{\pretopebs{#1}{#2}{#3}{#4}}}

\newcommand{\pretopecs}[5]{%
\begin{picture}(4,2.2)(-2,-0.2)%
\cell{-1.8}{1}{br}{#1}%
\cell{0}{2.2}{b}{#2}%
\cell{1.8}{1}{bl}{#3}%
\cell{0}{-0.2}{t}{#4}%
\cell{0}{0.8}{c}{#5}%
\put(-2,0){\line(1,2){1}}%
\put(-1,2){\line(1,0){2}}%
\put(1,2){\line(1,-2){1}}%
\put(-2,0){\line(1,0){4}}%
\end{picture}}

\mcm{\topecs}{5}{%
\ginitdims{4}{2.2}%
\sidespic{#1}%
\abovepic{#2}%
\sidespic{#3}%
\belowpic{#4}%
\present{\pretopecs{#1}{#2}{#3}{#4}{#5}}}

\newcommand{\pretopeds}[6]{%
\begin{picture}(4,2.5)(-2,-0.2)%
\cell{-2}{0.6}{br}{#1}%
\cell{-0.7}{2.2}{br}{#2}%
\cell{0.7}{2.2}{bl}{#3}%
\cell{2}{0.6}{bl}{#4}%
\cell{0}{-0.2}{t}{#5}%
\cell{0}{0.8}{c}{#6}%
\put(-2,0){\line(1,3){0.5}}%
\put(-1.5,1.5){\line(3,2){1.5}}%
\put(0,2.5){\line(3,-2){1.5}}%
\put(1.5,1.5){\line(1,-3){0.5}}%
\put(-2,0){\line(1,0){4}}%
\end{picture}}

\mcm{\topeds}{6}{%
\ginitdims{4}{2.5}%
\sidespic{#1}%
\abovepic{#2}%
\abovepic{#3}%
\sidespic{#4}%
\belowpic{#5}%
\present{\pretopeds{#1}{#2}{#3}{#4}{#5}{#6}}}

\newcommand{\pretopeqs}[5]{%
\begin{picture}(4,2.5)(-2,-0.2)%
\cell{-2}{0.6}{br}{#1}%
\cell{-1}{2.2}{br}{#2}%
\cell{2}{0.6}{bl}{#3}%
\cell{0}{-0.2}{t}{#4}%
\cell{0}{0.8}{c}{#5}%
\put(-2,0){\line(1,3){0.5}}%
\put(-1.5,1.5){\line(1,1){1}}%
\cell{0.9}{2.3}{c}{\ddots}
\put(1.5,1.5){\line(1,-3){0.5}}%
\put(-2,0){\line(1,0){4}}%
\end{picture}}

\mcm{\topeqs}{5}{%
\ginitdims{4}{2.5}%
\sidespic{#1}%
\abovepic{#2}%
\sidespic{#3}%
\belowpic{#4}%
\present{\pretopeqs{#1}{#2}{#3}{#4}{#5}}}

\newcommand{\pretopebases}[1]{%
\begin{picture}(4,0.4)(0,-0.2)%
\cell{2}{0.2}{b}{#1}%
\put(0,0){\line(1,0){4}}%
\end{picture}}

\mcm{\topebases}{1}{%
\ginitdims{4}{0.4}%
\abovepic{#1}%
\present{\pretopebases{#1}}}

\newcommand{\pregdots}[6]{%
\begin{picture}(5,8.4)(0,-2.7)%
\cell{2.5}{5.7}{b}{#1}%
\cell{1.5}{2.8}{b}{#2}%
\cell{1.5}{0.2}{t}{#3}%
\cell{2.5}{-2.7}{t}{#4}%
\cell{2.7}{4.25}{l}{#5}%
\cell{2.7}{-1.25}{l}{#6}%
\qbezier(0,1.5)(2.5,9.5)(5,1.5)%
\qbezier(0,1.5)(2.5,4)(5,1.5)%
\qbezier(0,1.5)(2.5,-1)(5,1.5)%
\qbezier(0,1.5)(2.5,-6.5)(5,1.5)%
\put(2.5,5.25){\vector(0,-1){2}}%
\put(2.5,-0.25){\vector(0,-1){2}}%
\cell{2.5}{1.7}{c}{\vdots}%
\put(5,1.5){\vector(1,-4){0}}%
\put(5,1.5){\vector(4,-3){0}}%
\put(5,1.5){\vector(4,3){0}}%
\put(5,1.5){\vector(1,4){0}}%
\end{picture}}

\mcm{\gdots}{6}{%
\ginitdims{5}{8.4}%
\abovepic{#1}%
\belowpic{#4}%
\present{\pregdots{#1}{#2}{#3}{#4}{#5}{#6}}}

\newlength{\volt}
\makeatletter

\def\diagram{\m@th\leftwidth=\z@ \rightwidth=\z@ \topheight=\z@
\botheight=\z@ \setbox\@picbox\hbox\bgroup}

\def\enddiagram{\egroup\wd\@picbox\rightwidth\unitlength
\ht\@picbox\topheight\unitlength \dp\@picbox\botheight\unitlength
\hskip\leftwidth\unitlength\box\@picbox}

\def\bfig{\begin{diagram}}
\def\efig{\end{diagram}}
\newcount\wideness \newcount\leftwidth \newcount\rightwidth
\newcount\highness \newcount\topheight \newcount\botheight

\def\ratchet#1#2{\ifnum#1<#2 \global #1=#2 \fi}

\def\putbox(#1,#2)#3{%
\horsize{\wideness}{#3} \divide\wideness by 2 {\advance\wideness
by #1 \ratchet{\rightwidth}{\wideness}} {\advance\wideness by -#1
\ratchet{\leftwidth}{\wideness}} \vertsize{\highness}{#3}
\divide\highness by 2 {\advance\highness by #2
\ratchet{\topheight}{\highness}} {\advance\highness by -#2
\ratchet{\botheight}{\highness}} \put(#1,#2){\makebox(0,0){$#3$}}}

\def\putlbox(#1,#2)#3{%
\horsize{\wideness}{#3} {\advance\wideness by #1
\ratchet{\rightwidth}{\wideness}} {\ratchet{\leftwidth}{-#1}}
\vertsize{\highness}{#3} \divide\highness by 2 {\advance\highness
by #2 \ratchet{\topheight}{\highness}} {\advance\highness by -#2
\ratchet{\botheight}{\highness}}
\put(#1,#2){\makebox(0,0)[l]{$#3$}}}

\def\putrbox(#1,#2)#3{%
\horsize{\wideness}{#3} {\ratchet{\rightwidth}{#1}}
{\advance\wideness by -#1 \ratchet{\leftwidth}{\wideness}}
\vertsize{\highness}{#3} \divide\highness by 2 {\advance\highness
by #2 \ratchet{\topheight}{\highness}} {\advance\highness by -#2
\ratchet{\botheight}{\highness}}
\put(#1,#2){\makebox(0,0)[r]{$#3$}}}

\def\adjust[#1]{} 

\newcount \coefa
\newcount \coefb
\newcount \coefc
\newcount\tempcounta
\newcount\tempcountb
\newcount\tempcountc
\newcount\tempcountd
\newcount\xext
\newcount\yext
\newcount\xoff
\newcount\yoff
\newcount\gap%
\newcount\arrowtypea
\newcount\arrowtypeb
\newcount\arrowtypec
\newcount\arrowtyped
\newcount\arrowtypee
\newcount\height
\newcount\width
\newcount\xpos
\newcount\ypos
\newcount\run
\newcount\rise
\newcount\arrowlength
\newcount\halflength
\newcount\arrowtype
\newdimen\tempdimen
\newdimen\xlen
\newdimen\ylen
\newsavebox{\tempboxa}%
\newsavebox{\tempboxb}%
\newsavebox{\tempboxc}%

\newdimen\w@dth

\def\setw@dth#1#2{\setbox\z@\hbox{\m@th$#1$}\w@dth=\wd\z@
\setbox\@ne\hbox{\m@th$#2$}\ifnum\w@dth<\wd\@ne \w@dth=\wd\@ne \fi
\advance\w@dth by 1.2em}

\def\t@^#1_#2{\allowbreak\def\n@one{#1}\def\n@two{#2}\mathrel
{\setw@dth{#1}{#2} \mathop{\hbox to
\w@dth{\rightarrowfill}}\limits \ifx\n@one\empty\else
^{\box\z@}\fi \ifx\n@two\empty\else _{\box\@ne}\fi}}
\def\t@@^#1{\@ifnextchar_{\t@^{#1}}{\t@^{#1}_{}}}
\def\to{\@ifnextchar^{\t@@}{\t@@^{}}}

\def\t@left^#1_#2{\def\n@one{#1}\def\n@two{#2}\mathrel{\setw@dth{#1}{#2}
\mathop{\hbox to \w@dth{\leftarrowfill}}\limits
\ifx\n@one\empty\else ^{\box\z@}\fi \ifx\n@two\empty\else
_{\box\@ne}\fi}}
\def\t@@left^#1{\@ifnextchar_{\t@left^{#1}}{\t@left^{#1}_{}}}
\def\toleft{\@ifnextchar^{\t@@left}{\t@@left^{}}}

\def\two@^#1_#2{\allowbreak
\def\n@one{#1}\def\n@two{#2}\mathrel{\setw@dth{#1}{#2}
\mathop{\vcenter{\lineskip\z@\baselineskip\z@
                 \hbox to \w@dth{\rightarrowfill}%
                 \hbox to \w@dth{\rightarrowfill}}%
       }\limits
\ifx\n@one\empty\else ^{\box\z@}\fi \ifx\n@two\empty\else
_{\box\@ne}\fi}}
\def\tw@@^#1{\@ifnextchar _{\two@^{#1}}{\two@^{#1}_{}}}
\def\two{\@ifnextchar ^{\tw@@}{\tw@@^{}}}

\def\tofr@^#1_#2{\def\n@one{#1}\def\n@two{#2}\mathrel{\setw@dth{#1}{#2}
\mathop{\vcenter{\hbox to \w@dth{\rightarrowfill}\kern-1.7ex
                 \hbox to \w@dth{\leftarrowfill}}%
       }\limits
\ifx\n@one\empty\else ^{\box\z@}\fi \ifx\n@two\empty\else
_{\box\@ne}\fi}}
\def\t@fr@^#1{\@ifnextchar_ {\tofr@^{#1}}{\tofr@^{#1}_{}}}
\def\tofro{\@ifnextchar^ {\t@fr@}{\t@fr@^{}}}

\def\mon{\mathop{\m@th\hbox to
      14.6\P@{\lasyb\char'51\hskip-2.1\P@$\arrext$\hss
$\mathord\rightarrow$}}\limits} 
\def\leftmono{\mathrel{\m@th\hbox to
14.6\P@{$\mathord\leftarrow$\hss$\arrext$\hskip-2.1\P@\lasyb\char'50%
}}\limits} 
\mathchardef\arrext="0200       

\def\settypes(#1,#2,#3){\arrowtypea#1 \arrowtypeb#2 \arrowtypec#3}
\def\settoheight#1#2{\setbox\@tempboxa\hbox{#2}#1\ht\@tempboxa\relax}%
\def\settodepth#1#2{\setbox\@tempboxa\hbox{#2}#1\dp\@tempboxa\relax}%
\def\settokens`#1`#2`#3`#4`{%
     \def\tokena{#1}\def\tokenb{#2}\def\tokenc{#3}\def\tokend{#4}}
\def\setsqparms[#1`#2`#3`#4;#5`#6]{%
\arrowtypea #1 \arrowtypeb #2 \arrowtypec #3 \arrowtyped #4
\width #5 \height #6 }
\def\setpos(#1,#2){\xpos=#1 \ypos#2}

\def\settriparms[#1`#2`#3;#4]{\settripairparms[#1`#2`#3`1`1;#4]}%

\def\settripairparms[#1`#2`#3`#4`#5;#6]{%
\arrowtypea #1 \arrowtypeb #2 \arrowtypec #3 \arrowtyped #4
\arrowtypee #5 \width #6 \height #6 }

\def\resetparms{\settripairparms[1`1`1`1`1;500]\width 500}

\resetparms

\def\mvector(#1,#2)#3{
\put(0,0){\vector(#1,#2){#3}}%
\put(0,0){\vector(#1,#2){26}}%
}
\def\evector(#1,#2)#3{{
\arrowlength #3
\put(0,0){\vector(#1,#2){\arrowlength}}%
\advance \arrowlength by-30
\put(0,0){\vector(#1,#2){\arrowlength}}%
}}

\def\horsize#1#2{%
\settowidth{\tempdimen}{$#2$}%
#1=\tempdimen \divide #1 by\unitlength }

\def\vertsize#1#2{%
\settoheight{\tempdimen}{$#2$}%
#1=\tempdimen
\settodepth{\tempdimen}{$#2$}%
\advance #1 by\tempdimen \divide #1 by\unitlength }

\def\putvector(#1,#2)(#3,#4)#5#6{{%
\ifnum3<\arrowtype \putdashvector(#1,#2)(#3,#4)#5\arrowtype \else
\ifnum\arrowtype<-3 \putdashvector(#1,#2)(#3,#4)#5\arrowtype \else
\xpos=#1 \ypos=#2 \run=#3 \rise=#4 \arrowlength=#5 \ifnum
\arrowtype<0
    \ifnum \run=0
        \advance \ypos by-\arrowlength
    \else
        \tempcounta \arrowlength
        \multiply \tempcounta by\rise
        \divide \tempcounta by\run
        \ifnum\run>0
            \advance \xpos by\arrowlength
            \advance \ypos by\tempcounta
        \else
            \advance \xpos by-\arrowlength
            \advance \ypos by-\tempcounta
        \fi
    \fi
    \multiply \arrowtype by-1
    \multiply \rise by-1
    \multiply \run by-1
\fi \ifcase \arrowtype
\or \put(\xpos,\ypos){\vector(\run,\rise){\arrowlength}}%
\or \put(\xpos,\ypos){\mvector(\run,\rise)\arrowlength}%
\or \put(\xpos,\ypos){\evector(\run,\rise){\arrowlength}}%
\fi\fi\fi }}

\def\putsplitvector(#1,#2)#3#4{
\xpos #1 \ypos #2 \arrowtype #4 \halflength #3 \arrowlength #3
\gap 140 \advance \halflength by-\gap \divide \halflength by2
\ifnum\arrowtype>0
   \ifcase \arrowtype
   \or \put(\xpos,\ypos){\line(0,-1){\halflength}}%
       \advance\ypos by-\halflength
       \advance\ypos by-\gap
       \put(\xpos,\ypos){\vector(0,-1){\halflength}}%
   \or \put(\xpos,\ypos){\line(0,-1)\halflength}%
       \put(\xpos,\ypos){\vector(0,-1)3}%
       \advance\ypos by-\halflength
       \advance\ypos by-\gap
       \put(\xpos,\ypos){\vector(0,-1){\halflength}}%
   \or \put(\xpos,\ypos){\line(0,-1)\halflength}%
       \advance\ypos by-\halflength
       \advance\ypos by-\gap
       \put(\xpos,\ypos){\evector(0,-1){\halflength}}%
   \fi
\else \arrowtype=-\arrowtype
   \ifcase\arrowtype
   \or \advance \ypos by-\arrowlength
       \put(\xpos,\ypos){\line(0,1){\halflength}}%
       \advance\ypos by\halflength
       \advance\ypos by\gap
       \put(\xpos,\ypos){\vector(0,1){\halflength}}%
   \or \advance \ypos by-\arrowlength
       \put(\xpos,\ypos){\line(0,1)\halflength}%
       \put(\xpos,\ypos){\vector(0,1)3}%
       \advance\ypos by\halflength
       \advance\ypos by\gap
       \put(\xpos,\ypos){\vector(0,1){\halflength}}%
   \or \advance \ypos by-\arrowlength
       \put(\xpos,\ypos){\line(0,1)\halflength}%
       \advance\ypos by\halflength
       \advance\ypos by\gap
       \put(\xpos,\ypos){\evector(0,1){\halflength}}%
   \fi
\fi }

\def\putmorphism(#1)(#2,#3)[#4`#5`#6]#7#8#9{{%
\run #2 \rise #3 \ifnum\rise=0
  \puthmorphism(#1)[#4`#5`#6]{#7}{#8}#9%
\else\ifnum\run=0
  \putvmorphism(#1)[#4`#5`#6]{#7}{#8}#9%
\else
\setpos(#1)%
\arrowlength #7 \arrowtype #8 \ifnum\run=0 \else\ifnum\rise=0
\else \ifnum\run>0
    \coefa=1
\else
   \coefa=-1
\fi \ifnum\arrowtype>0
   \coefb=0
   \coefc=-1
\else
   \coefb=\coefa
   \coefc=1
   \arrowtype=-\arrowtype
\fi \width=2 \multiply \width by\run \divide \width by\rise
\ifnum \width<0  \width=-\width\fi \advance\width by60 \if l#9
\width=-\width\fi
\putbox(\xpos,\ypos){#4}
{\multiply \coefa by\arrowlength
\advance\xpos by\coefa \multiply \coefa by\rise \divide \coefa
by\run \advance \ypos by\coefa
\putbox(\xpos,\ypos){#5} }%
{\multiply \coefa by\arrowlength
\divide \coefa by2 \advance \xpos by\coefa \advance \xpos by\width
\multiply \coefa by\rise \divide \coefa by\run \advance \ypos
by\coefa
\if l#9%
   \putrbox(\xpos,\ypos){#6}%
\else\if r#9%
   \putlbox(\xpos,\ypos){#6}%
\fi\fi }%
{\multiply \rise by-\coefc
\multiply \run by-\coefc \multiply \coefb by\arrowlength \advance
\xpos by\coefb \multiply \coefb by\rise \divide \coefb by\run
\advance \ypos by\coefb \multiply \coefc by70 \advance \ypos
by\coefc \multiply \coefc by\run \divide \coefc by\rise \advance
\xpos by\coefc \multiply \coefa by140 \multiply \coefa by\run
\divide \coefa by\rise \advance \arrowlength by\coefa
\ifcase\arrowtype
\or \put(\xpos,\ypos){\vector(\run,\rise){\arrowlength}}%
\or \put(\xpos,\ypos){\mvector(\run,\rise){\arrowlength}}%
\or \put(\xpos,\ypos){\evector(\run,\rise){\arrowlength}}%
\fi}\fi\fi\fi\fi}}

\newcount\numbdashes \newcount\lengthdash \newcount\increment

\def\howmanydashes{
\numbdashes=\arrowlength \lengthdash=40 \divide\numbdashes by
\lengthdash \lengthdash=\arrowlength \divide\lengthdash by
\numbdashes
\increment=\lengthdash \multiply\lengthdash by 3
\divide\lengthdash by 5 }

\def\putdashvector(#1)(#2,#3)#4#5{%
\ifnum#3=0 \putdashhvector(#1){#4}#5 \else \ifnum#2=0
\putdashvvector(#1){#4}#5\fi\fi}

\def\putdashhvector(#1,#2)#3#4{{%
\arrowlength=#3 \howmanydashes
\multiput(#1,#2)(\increment,0){\numbdashes}%
{\vrule height .4pt width \lengthdash\unitlength} \arrowtype=#4
\xpos=#1 \ifnum\arrowtype<0 \advance\arrowtype by 7 \fi
\ifcase\arrowtype \or \advance\xpos by 10
    \put(\xpos,#2){\vector(-1,0){\lengthdash}}
    \advance\xpos by 40
    \put(\xpos,#2){\vector(-1,0){\lengthdash}}
\or \advance \xpos by 10
    \put(\xpos,#2){\vector(-1,0){\lengthdash}}
    \advance\xpos by  \arrowlength
    \advance\xpos by  -50
    \put(\xpos,#2){\vector(-1,0){\lengthdash}}
\or \advance\xpos by 10
    \put(\xpos,#2){\vector(-1,0){\lengthdash}}
\or \advance\xpos by \arrowlength
    \advance\xpos by -\lengthdash
    \put(\xpos,#2){\vector(1,0){\lengthdash}}
\or {\advance\xpos by 10
    \put(\xpos,#2){\vector(1,0){\lengthdash}}}
    \advance\xpos by \arrowlength
    \advance\xpos by -\lengthdash
    \put(\xpos,#2){\vector(1,0){\lengthdash}}
\or \advance\xpos by \arrowlength
    \advance\xpos by -\lengthdash
    \put(\xpos,#2){\vector(1,0){\lengthdash}}
    \advance\xpos by -40
    \put(\xpos,#2){\vector(1,0){\lengthdash}}
   \fi
}}

\def\putdashvvector(#1,#2)#3#4{{%
\arrowlength=#3 \howmanydashes \ypos=#2 \advance\ypos by
-\arrowlength
\multiput(#1,#2)(0,\increment){\numbdashes}%
    {\vrule width .4pt height \lengthdash\unitlength}
\arrowtype=#4 \ypos=#2 \ifnum\arrowtype<0 \advance\arrowtype by 7
\fi \ifcase\arrowtype \or \advance\ypos by \arrowlength
\advance\ypos by -40
    \put(#1,\ypos){\vector(0,1){\lengthdash}}
    \advance\ypos by -40
    \put(#1,\ypos){\vector(0,1){\lengthdash}}
\or \advance\ypos by 10
    \put(#1,\ypos){\vector(0,1){\lengthdash}}
    \advance\ypos by \arrowlength \advance\ypos by -40
    \put(#1,\ypos){\vector(0,1){\lengthdash}}
\or \advance\ypos by \arrowlength \advance\ypos by -40
    \put(#1,\ypos){\vector(0,1){\lengthdash}}
\or \advance\ypos by 10
    \put(#1,\ypos){\vector(0,-1){\lengthdash}}
\or \advance\ypos by 10
    \put(#1,\ypos){\vector(0,-1){\lengthdash}}
    \advance\ypos by \arrowlength \advance\ypos by -40
    \put(#1,\ypos){\vector(0,-1){\lengthdash}}
\or \advance\ypos by 10
    \put(#1,\ypos){\vector(0,-1){\lengthdash}}
    \advance\ypos by 40
    \put(#1,\ypos){\vector(0,-1){\lengthdash}}
\fi }}

\def\puthmorphism(#1,#2)[#3`#4`#5]#6#7#8{{%
\xpos #1 \ypos #2 \width #6 \arrowlength #6 \arrowtype=#7
\putbox(\xpos,\ypos){#3\vphantom{#4}}%
{\advance \xpos by\arrowlength
\putbox(\xpos,\ypos){\vphantom{#3}#4}}%
\horsize{\tempcounta}{#3}%
\horsize{\tempcountb}{#4}%
\divide \tempcounta by2 \divide \tempcountb by2 \advance
\tempcounta by30 \advance \tempcountb by30 \advance \xpos
by\tempcounta \advance \arrowlength by-\tempcounta \advance
\arrowlength by-\tempcountb
\putvector(\xpos,\ypos)(1,0)\arrowlength\arrowtype \divide
\arrowlength by2 \advance \xpos by\arrowlength
\vertsize{\tempcounta}{#5}%
\divide\tempcounta by2 \advance \tempcounta by20
\if a#8 %
   \advance \ypos by\tempcounta
   \putbox(\xpos,\ypos){#5}%
\else
   \advance \ypos by-\tempcounta
   \putbox(\xpos,\ypos){#5}%
\fi}}

\def\putvmorphism(#1,#2)[#3`#4`#5]#6#7#8{{%
\xpos #1 \ypos #2 \arrowlength #6 \arrowtype #7
\settowidth{\xlen}{$#5$}%
\putbox(\xpos,\ypos){#3}%
{\advance \ypos by-\arrowlength
\putbox(\xpos,\ypos){#4}}%
{\advance\arrowlength by-140 \advance \ypos by-70 \ifdim\xlen>0pt
   \if m#8%
      \putsplitvector(\xpos,\ypos)\arrowlength\arrowtype
   \else
   \putvector(\xpos,\ypos)(0,-1)\arrowlength\arrowtype
   \fi
\else
   \putvector(\xpos,\ypos)(0,-1)\arrowlength\arrowtype
\fi}%
\ifdim\xlen>0pt
   \divide \arrowlength by2
   \advance\ypos by-\arrowlength
   \if l#8%
      \advance \xpos by-40
      \putrbox(\xpos,\ypos){#5}%
   \else\if r#8%
      \advance \xpos by40
      \putlbox(\xpos,\ypos){#5}%
   \else
      \putbox(\xpos,\ypos){#5}%
   \fi\fi
\fi }}

\def\putsquarep<#1>(#2)[#3;#4`#5`#6`#7]{{%
\setsqparms[#1]%
\setpos(#2)%
\settokens`#3`%
\puthmorphism(\xpos,\ypos)[\tokenc`\tokend`{#7}]{\width}{\arrowtyped}b%
\advance\ypos by \height
\puthmorphism(\xpos,\ypos)[\tokena`\tokenb`{#4}]{\width}{\arrowtypea}a%
\putvmorphism(\xpos,\ypos)[``{#5}]{\height}{\arrowtypeb}l%
\advance\xpos by \width
\putvmorphism(\xpos,\ypos)[``{#6}]{\height}{\arrowtypec}r%
}}

\def\putsquare{\@ifnextchar <{\putsquarep}{\putsquarep%
   <\arrowtypea`\arrowtypeb`\arrowtypec`\arrowtyped;\width`\height>}}
\def\square{\@ifnextchar< {\squarep}{\squarep
   <\arrowtypea`\arrowtypeb`\arrowtypec`\arrowtyped;\width`\height>}}
\def\squarep<#1>[#2`#3`#4`#5;#6`#7`#8`#9]{{
\setsqparms[#1]
\diagram
\putsquarep<\arrowtypea`\arrowtypeb`\arrowtypec`
\arrowtyped;\width`\height>
(0,0)[#2`#3`#4`{#5};#6`#7`#8`{#9}]
\enddiagram
}}                                                 
\def\putptrianglep<#1>(#2,#3)[#4`#5`#6;#7`#8`#9]{{%
\settriparms[#1]%
\xpos=#2 \ypos=#3 \advance\ypos by \height
\puthmorphism(\xpos,\ypos)[#4`#5`{#7}]{\height}{\arrowtypea}a%
\putvmorphism(\xpos,\ypos)[`#6`{#8}]{\height}{\arrowtypeb}l%
\advance\xpos by\height
\putmorphism(\xpos,\ypos)(-1,-1)[``{#9}]{\height}{\arrowtypec}r%
}}

\def\putptriangle{\@ifnextchar <{\putptrianglep}{\putptrianglep
   <\arrowtypea`\arrowtypeb`\arrowtypec;\height>}}
\def\ptriangle{\@ifnextchar <{\ptrianglep}{\ptrianglep
   <\arrowtypea`\arrowtypeb`\arrowtypec;\height>}}
\def\ptrianglep<#1>[#2`#3`#4;#5`#6`#7]{{
\settriparms[#1]
\diagram
\putptrianglep<\arrowtypea`\arrowtypeb`
\arrowtypec;\height>
(0,0)[#2`#3`#4;#5`#6`{#7}]
\enddiagram
}}                                            

\def\putqtrianglep<#1>(#2,#3)[#4`#5`#6;#7`#8`#9]{{%
\settriparms[#1]%
\xpos=#2 \ypos=#3 \advance\ypos by\height
\puthmorphism(\xpos,\ypos)[#4`#5`{#7}]{\height}{\arrowtypea}a%
\putmorphism(\xpos,\ypos)(1,-1)[``{#8}]{\height}{\arrowtypeb}l%
\advance\xpos by\height
\putvmorphism(\xpos,\ypos)[`#6`{#9}]{\height}{\arrowtypec}r%
}}

\def\putqtriangle{\@ifnextchar <{\putqtrianglep}{\putqtrianglep
   <\arrowtypea`\arrowtypeb`\arrowtypec;\height>}}
\def\qtriangle{\@ifnextchar <{\qtrianglep}{\qtrianglep
   <\arrowtypea`\arrowtypeb`\arrowtypec;\height>}}
\def\qtrianglep<#1>[#2`#3`#4;#5`#6`#7]{{
\settriparms[#1]
\width=\height                                
\diagram
\putqtrianglep<\arrowtypea`\arrowtypeb`
\arrowtypec;\height>
(0,0)[#2`#3`#4;#5`#6`{#7}]
\enddiagram
}}

\def\putdtrianglep<#1>(#2,#3)[#4`#5`#6;#7`#8`#9]{{%
\settriparms[#1]%
\xpos=#2 \ypos=#3
\puthmorphism(\xpos,\ypos)[#5`#6`{#9}]{\height}{\arrowtypec}b%
\advance\xpos by \height \advance\ypos by\height
\putmorphism(\xpos,\ypos)(-1,-1)[``{#7}]{\height}{\arrowtypea}l%
\putvmorphism(\xpos,\ypos)[#4``{#8}]{\height}{\arrowtypeb}r%
}}

\def\putdtriangle{\@ifnextchar <{\putdtrianglep}{\putdtrianglep
   <\arrowtypea`\arrowtypeb`\arrowtypec;\height>}}
\def\dtriangle{\@ifnextchar <{\dtrianglep}{\dtrianglep
   <\arrowtypea`\arrowtypeb`\arrowtypec;\height>}}
\def\dtrianglep<#1>[#2`#3`#4;#5`#6`#7]{{
\settriparms[#1]
\width=\height                                
\diagram
\putdtrianglep<\arrowtypea`\arrowtypeb`
\arrowtypec;\height>
(0,0)[#2`#3`#4;#5`#6`{#7}]
\enddiagram
}}

\def\putbtrianglep<#1>(#2,#3)[#4`#5`#6;#7`#8`#9]{{%
\settriparms[#1]%
\xpos=#2 \ypos=#3
\puthmorphism(\xpos,\ypos)[#5`#6`{#9}]{\height}{\arrowtypec}b%
\advance\ypos by\height
\putmorphism(\xpos,\ypos)(1,-1)[``{#8}]{\height}{\arrowtypeb}r%
\putvmorphism(\xpos,\ypos)[#4``{#7}]{\height}{\arrowtypea}l%
}}

\def\putbtriangle{\@ifnextchar <{\putbtrianglep}{\putbtrianglep
   <\arrowtypea`\arrowtypeb`\arrowtypec;\height>}}
\def\btriangle{\@ifnextchar <{\btrianglep}{\btrianglep
   <\arrowtypea`\arrowtypeb`\arrowtypec;\height>}}
\def\btrianglep<#1>[#2`#3`#4;#5`#6`#7]{{
\settriparms[#1]
\width=\height                               
\diagram
\putbtrianglep<\arrowtypea`\arrowtypeb`
\arrowtypec;\height>
(0,0)[#2`#3`#4;#5`#6`{#7}]
\enddiagram
}}

\def\putAtrianglep<#1>(#2,#3)[#4`#5`#6;#7`#8`#9]{{%
\settriparms[#1]%
\xpos=#2 \ypos=#3 {\multiply \height by2
\puthmorphism(\xpos,\ypos)[#5`#6`{#9}]{\height}{\arrowtypec}b}%
\advance\xpos by\height \advance\ypos by\height
\putmorphism(\xpos,\ypos)(-1,-1)[#4``{#7}]{\height}{\arrowtypea}l%
\putmorphism(\xpos,\ypos)(1,-1)[``{#8}]{\height}{\arrowtypeb}r%
}}

\def\putAtriangle{\@ifnextchar <{\putAtrianglep}{\putAtrianglep
   <\arrowtypea`\arrowtypeb`\arrowtypec;\height>}}
\def\Atriangle{\@ifnextchar <{\Atrianglep}{\Atrianglep
   <\arrowtypea`\arrowtypeb`\arrowtypec;\height>}}
\def\Atrianglep<#1>[#2`#3`#4;#5`#6`#7]{{
\settriparms[#1]
\width=\height                                     
\diagram
\putAtrianglep<\arrowtypea`\arrowtypeb`
\arrowtypec;\height>
(0,0)[#2`#3`#4;#5`#6`{#7}]
\enddiagram
}}

\def\putAtrianglepairp<#1>(#2)[#3;#4`#5`#6`#7`#8]{{%
\settripairparms[#1]%
\setpos(#2)%
\settokens`#3`%
\puthmorphism(\xpos,\ypos)[\tokenb`\tokenc`{#7}]{\height}{\arrowtyped}b%
\advance\xpos by\height
\puthmorphism(\xpos,\ypos)[\phantom{\tokenc}`\tokend`{#8}]%
{\height}{\arrowtypee}b%
\advance\ypos by\height
\putmorphism(\xpos,\ypos)(-1,-1)[\tokena``{#4}]{\height}{\arrowtypea}l%
\putvmorphism(\xpos,\ypos)[``{#5}]{\height}{\arrowtypeb}m%
\putmorphism(\xpos,\ypos)(1,-1)[``{#6}]{\height}{\arrowtypec}r%
}}

\def\putAtrianglepair{\@ifnextchar <{\putAtrianglepairp}{\putAtrianglepairp%
   <\arrowtypea`\arrowtypeb`\arrowtypec`\arrowtyped`\arrowtypee;\height>}}
\def\Atrianglepair{\@ifnextchar <{\Atrianglepairp}{\Atrianglepairp%
   <\arrowtypea`\arrowtypeb`\arrowtypec`\arrowtyped`\arrowtypee;\height>}}

\def\Atrianglepairp<#1>[#2;#3`#4`#5`#6`#7]{{
\settripairparms[#1]
\settokens`#2`
\width=\height                                
\diagram
\putAtrianglepairp                            
<\arrowtypea`\arrowtypeb`\arrowtypec`
\arrowtyped`\arrowtypee;\height>
(0,0)[{#2};#3`#4`#5`#6`{#7}]
\enddiagram
}}

\def\putVtrianglep<#1>(#2,#3)[#4`#5`#6;#7`#8`#9]{{%
\settriparms[#1]%
\xpos=#2 \ypos=#3 \advance\ypos by\height {\multiply\height by2
\puthmorphism(\xpos,\ypos)[#4`#5`{#7}]{\height}{\arrowtypea}a}%
\putmorphism(\xpos,\ypos)(1,-1)[`#6`{#8}]{\height}{\arrowtypeb}l%
\advance\xpos by\height \advance\xpos by\height
\putmorphism(\xpos,\ypos)(-1,-1)[``{#9}]{\height}{\arrowtypec}r%
}}

\def\putVtriangle{\@ifnextchar <{\putVtrianglep}{\putVtrianglep
   <\arrowtypea`\arrowtypeb`\arrowtypec;\height>}}
\def\Vtriangle{\@ifnextchar <{\Vtrianglep}{\Vtrianglep
   <\arrowtypea`\arrowtypeb`\arrowtypec;\height>}}
\def\Vtrianglep<#1>[#2`#3`#4;#5`#6`#7]{{
\settriparms[#1]
\width=\height                                 
\diagram
\putVtrianglep<\arrowtypea`\arrowtypeb`
\arrowtypec;\height>
(0,0)[#2`#3`#4;#5`#6`{#7}]
\enddiagram
}}

\def\putVtrianglepairp<#1>(#2)[#3;#4`#5`#6`#7`#8]{{
\settripairparms[#1]%
\setpos(#2)%
\settokens`#3`%
\advance\ypos by\height
\putmorphism(\xpos,\ypos)(1,-1)[`\tokend`{#6}]{\height}{\arrowtypec}l%
\puthmorphism(\xpos,\ypos)[\tokena`\tokenb`{#4}]{\height}{\arrowtypea}a%
\advance\xpos by\height
\puthmorphism(\xpos,\ypos)[\phantom{\tokenb}`\tokenc`{#5}]%
{\height}{\arrowtypeb}a%
\putvmorphism(\xpos,\ypos)[``{#7}]{\height}{\arrowtyped}m%
\advance\xpos by\height
\putmorphism(\xpos,\ypos)(-1,-1)[``{#8}]{\height}{\arrowtypee}r%
}}

\def\putVtrianglepair{\@ifnextchar <{\putVtrianglepairp}{\putVtrianglepairp%
    <\arrowtypea`\arrowtypeb`\arrowtypec`\arrowtyped`\arrowtypee;\height>}}
\def\Vtrianglepair{\@ifnextchar <{\Vtrianglepairp}{\Vtrianglepairp%
    <\arrowtypea`\arrowtypeb`\arrowtypec`\arrowtyped`\arrowtypee;\height>}}
\def\Vtrianglepairp<#1>[#2;#3`#4`#5`#6`#7]{{
\settripairparms[#1]
\settokens`#2`
\diagram
\putVtrianglepairp                             
<\arrowtypea`\arrowtypeb`\arrowtypec`
\arrowtyped`\arrowtypee;\height>
(0,0)[{#2};#3`#4`#5`#6`{#7}]
\enddiagram
}}

\def\putCtrianglep<#1>(#2,#3)[#4`#5`#6;#7`#8`#9]{{%
\settriparms[#1]%
\xpos=#2 \ypos=#3 \advance\ypos by\height
\putmorphism(\xpos,\ypos)(1,-1)[``{#9}]{\height}{\arrowtypec}l%
\advance\xpos by\height \advance\ypos by\height
\putmorphism(\xpos,\ypos)(-1,-1)[#4`#5`{#7}]{\height}{\arrowtypea}l%
{\multiply\height by 2
\putvmorphism(\xpos,\ypos)[`#6`{#8}]{\height}{\arrowtypeb}r}%
}}

\def\putCtriangle{\@ifnextchar <{\putCtrianglep}{\putCtrianglep
    <\arrowtypea`\arrowtypeb`\arrowtypec;\height>}}
\def\Ctriangle{\@ifnextchar <{\Ctrianglep}{\Ctrianglep
    <\arrowtypea`\arrowtypeb`\arrowtypec;\height>}}
\def\Ctrianglep<#1>[#2`#3`#4;#5`#6`#7]{{
\settriparms[#1]
\width=\height                               
\diagram
\putCtrianglep<\arrowtypea`\arrowtypeb`
\arrowtypec;\height>
(0,0)[#2`#3`#4;#5`#6`{#7}]
\enddiagram
}}                                           
\def\putDtrianglep<#1>(#2,#3)[#4`#5`#6;#7`#8`#9]{{%
\settriparms[#1]%
\xpos=#2 \ypos=#3 \advance\xpos by\height \advance\ypos by\height
\putmorphism(\xpos,\ypos)(-1,-1)[``{#9}]{\height}{\arrowtypec}r%
\advance\xpos by-\height \advance\ypos by\height
\putmorphism(\xpos,\ypos)(1,-1)[`#5`{#8}]{\height}{\arrowtypeb}r%
{\multiply\height by 2
\putvmorphism(\xpos,\ypos)[#4`#6`{#7}]{\height}{\arrowtypea}l}%
}}

\def\putDtriangle{\@ifnextchar <{\putDtrianglep}{\putDtrianglep
    <\arrowtypea`\arrowtypeb`\arrowtypec;\height>}}
\def\Dtriangle{\@ifnextchar <{\Dtrianglep}{\Dtrianglep
   <\arrowtypea`\arrowtypeb`\arrowtypec;\height>}}
\def\Dtrianglep<#1>[#2`#3`#4;#5`#6`#7]{{
\settriparms[#1]
\width=\height                              
\diagram
\putDtrianglep<\arrowtypea`\arrowtypeb`
\arrowtypec;\height>
(0,0)[#2`#3`#4;#5`#6`{#7}]
\enddiagram
}}                                          
\def\setrecparms[#1`#2]{\width=#1 \height=#2}%

\def\recursep<#1`#2>[#3;#4`#5`#6`#7`#8]{{\m@th
\width=#1 \height=#2 \settokens`#3`
\settowidth{\tempdimen}{$\tokena$} \ifdim\tempdimen=0pt
  \savebox{\tempboxa}{\hbox{$\tokenb$}}%
  \savebox{\tempboxb}{\hbox{$\tokend$}}%
  \savebox{\tempboxc}{\hbox{$#6$}}%
\else
  \savebox{\tempboxa}{\hbox{$\hbox{$\tokena$}\times\hbox{$\tokenb$}$}}%
  \savebox{\tempboxb}{\hbox{$\hbox{$\tokena$}\times\hbox{$\tokend$}$}}%
  \savebox{\tempboxc}{\hbox{$\hbox{$\tokena$}\times\hbox{$#6$}$}}%
\fi \ypos=\height \divide\ypos by 2 \xpos=\ypos \advance\xpos by
\width \bfig
\putCtrianglep<-1`1`1;\ypos>(0,0)[`\tokenc`;#5`#6`{#7}]%
\puthmorphism(\ypos,0)[\tokend`\usebox{\tempboxb}`{#8}]{\width}{-1}b%
\puthmorphism(\ypos,\height)[\tokenb`\usebox{\tempboxa}`{#4}]{\width}{-1}a%
\advance\ypos by \width
\putvmorphism(\ypos,\height)[``\usebox{\tempboxc}]{\height}1r%
\efig }}

\def\recurse{\@ifnextchar <{\recursep}{\recursep<\width`\height>}}

\def\puttwohmorphisms(#1,#2)[#3`#4;#5`#6]#7#8#9{{%
\puthmorphism(#1,#2)[#3`#4`]{#7}0a \ypos=#2 \advance\ypos by 20
\puthmorphism(#1,\ypos)[\phantom{#3}`\phantom{#4}`#5]{#7}{#8}a
\advance\ypos by -40
\puthmorphism(#1,\ypos)[\phantom{#3}`\phantom{#4}`#6]{#7}{#9}b }}

\def\puttwovmorphisms(#1,#2)[#3`#4;#5`#6]#7#8#9{{%
\putvmorphism(#1,#2)[#3`#4`]{#7}0a \xpos=#1 \advance\xpos by -20
\putvmorphism(\xpos,#2)[\phantom{#3}`\phantom{#4}`#5]{#7}{#8}l
\advance\xpos by 40
\putvmorphism(\xpos,#2)[\phantom{#3}`\phantom{#4}`#6]{#7}{#9}r }}

\def\puthcoequalizer(#1)[#2`#3`#4;#5`#6`#7]#8#9{{%
\setpos(#1)%
\puttwohmorphisms(\xpos,\ypos)[#2`#3;#5`#6]{#8}11%
\advance\xpos by #8
\puthmorphism(\xpos,\ypos)[\phantom{#3}`#4`#7]{#8}1{#9} }}

\def\putvcoequalizer(#1)[#2`#3`#4;#5`#6`#7]#8#9{{%
\setpos(#1)%
\puttwovmorphisms(\xpos,\ypos)[#2`#3;#5`#6]{#8}11%
\advance\ypos by -#8
\putvmorphism(\xpos,\ypos)[\phantom{#3}`#4`#7]{#8}1{#9} }}

\def\putthreehmorphisms(#1)[#2`#3;#4`#5`#6]#7(#8)#9{{%
\setpos(#1) \settypes(#8)
\if a#9 %
     \vertsize{\tempcounta}{#5}%
     \vertsize{\tempcountb}{#6}%
     \ifnum \tempcounta<\tempcountb \tempcounta=\tempcountb \fi
\else
     \vertsize{\tempcounta}{#4}%
     \vertsize{\tempcountb}{#5}%
     \ifnum \tempcounta<\tempcountb \tempcounta=\tempcountb \fi
\fi \advance \tempcounta by 60
\puthmorphism(\xpos,\ypos)[#2`#3`#5]{#7}{\arrowtypeb}{#9}
\advance\ypos by \tempcounta
\puthmorphism(\xpos,\ypos)[\phantom{#2}`\phantom{#3}`#4]{#7}{\arrowtypea}{#9}
\advance\ypos by -\tempcounta \advance\ypos by -\tempcounta
\puthmorphism(\xpos,\ypos)[\phantom{#2}`\phantom{#3}`#6]{#7}{\arrowtypec}{#9}
}}

\def\setarrowtoks[#1`#2`#3`#4`#5`#6]{%
\def\toka{#1}
\def\tokb{#2}
\def\tokc{#3}
\def\tokd{#4}
\def\toke{#5}
\def\tokf{#6}
}
\def\hex{\@ifnextchar <{\hexp}{\hexp<1000`400>}}
\def\hexp<#1`#2>[#3`#4`#5`#6`#7`#8;#9]{%
\setarrowtoks[#9] \yext=#2 \advance \yext by #2 \xext=#1
\advance\xext by \yext \bfig
\putCtriangle<-1`0`1;#2>(0,0)[`#5`;\tokb``\tokd] \xext=#1
\yext=#2 \advance \yext by #2
\putsquare<1`0`0`1;\xext`\yext>(#2,0)[#3`#4`#7`#8;\toka```\tokf]
\advance \xext by #2
\putDtriangle<0`1`-1;#2>(\xext,0)[`#6`;`\tokc`\toke] \efig }

\makeatother

\input{tcilatex}

\begin{document}

\title{Jet--Ricci Geometry of Time-Dependent Human Biomechanics}\author{Tijana T. Ivancevic\\ {\small Society for Nonlinear Dynamics in Human Factors, Adelaide, Australia}\\
{\small and}\\
{\small CITECH Research IP Pty Ltd, Adelaide, Australia}\\
{\small e-mail: ~tijana.ivancevic@alumni.adelaide.edu.au}}\date{}\maketitle

\begin{abstract}
We propose the time-dependent generalization of an `ordinary' autonomous human biomechanics, in which \emph{total mechanical + biochemical energy is not conserved}. We introduce a general framework for time-dependent biomechanics in terms of jet manifolds derived from the extended musculo-skeletal configuration manifold. The corresponding Riemannian geometrical evolution follows the Ricci flow diffusion. In particular, we show that the exponential-like decay of total biomechanical energy (due to exhaustion of biochemical resources) is closely related to the Ricci flow\footnote{Ricci flow is a current hot topic in pure mathematics for which the latest Fields Medal was awarded to G. Perelman for the proof of 100 year-old Poincar\'e Conjecture.} on the biomechanical configuration manifold.\\

\end{abstract}


\section{Introduction}

It is a well-known fact that most of Hamiltonian and Lagrangian dynamics in physics are based on \emph{assumption of a total energy conservation}. Even more, the word ``Hamiltonian" usually means ``conservative", while Lagrangian formalism is usually proved to be equivalent to Hamiltonian (therefore also conservative), as derived from a conservative Lagrangian energy function (for a comprehensive review see, e.g. \cite{GaneshADG}). The straightforward application of Hamiltonian/Lagrangian formalisms to human biomechanics would naturally inherit this conservative assumption.\footnote{An engineering-type approach to this problem would be simply adding dissipation and forcing into equations of motion, without deriving them form some ``dissipative Lagrangian" and/or ``dissipative Hamiltonian"; this yields an extended Hamiltonian and/or Lagrangian formalism (see \cite{GaneshSprSml,GaneshWSc,GaneshSprBig,StrAttr,TijIJHR,TijNis,TijNL,TijSpr}).}

And this works fine for most individual movement simulations and predictions, in which the total human energy dissipations are insignificant. However, if we analyze a 100\,m-dash sprinting motion, which is in case of top athletes finished under 10\,s, we can recognize a significant slow-down after about 70\,m in \emph{all} athletes -- despite of their strong intention to finish and win the race, which is an obvious sign of the total energy dissipation. This can be
seen, for example, in a current record-braking speed–distance curve of Usain Bolt, the  world-record holder with 9.69 s \cite{SciSport}, or in a former record-braking speed–distance curve of Carl Lewis, the former world-record holder (and 9 time Olympic gold medalist) with 9.86 s (see Figure 3.7 in \cite{TijSpr}). In other words, the \emph{total mechanical + biochemical energy} of a sprinter \emph{cannot be conserved} even for 10\,s. So, if we want to develop a realistic model of intensive human motion that is longer than 7--8\,s (not to speak for instance of a 4 hour tennis match), we necessarily need to use the more advanced formalism of time-dependent mechanics.

Similarly, if we analyze individual movements of gymnasts or pirouettes in ice skating, we can clearly see that the high speed of these movements is based on quickly-varying mass-inertia distribution of various body segments (mostly arms and legs). As the total mass-inertia matrix of a biomechanical system corresponds to the Riemannian metric tensor of its configuration manifold, we can formulate this problem in terms of time-dependent Riemannian geometry \cite{GaneshSprBig,GaneshADG}.

The purpose of this paper is to introduce a general framework for time-dependent biomechanics, consisting of:
\begin{enumerate}
  \item human biomechanical configuration manifold and its (co)tangent bundles;
  \item biomechanical jet spaces and prolongation of locomotion vector-fields developed on the biomechanical configuration manifold; and
  \item time-dependent Lagrangian dynamics using biomechanical jet spaces.
\end{enumerate}
In addition, we will show that Riemannian geometrical basis of this framework is defined by the Ricci flow. In particular, we will show that the exponential-like decay of total biomechanical energy (due to exhaustion of biochemical resources \cite{TijSpr}) is closely related to the Ricci flow on the configuration manifold of human motion.

\section{Human Biomechanical Manifold and its (Co)Tangent Bundles}

\subsection{Humanoid Robot Dynamics}

Recall from \cite{TijIJHR} that representation of an ideal humanoid--robot motion is rigorously defined in
terms of {rotational} constrained $SO(3)$--groups in all main robot joints. Therefore, the {configuration manifold}
$Q_{rob}$
for humanoid dynamics is defined as a topological product of all included $%
SO(3)$ groups, $Q_{rob}=\prod_{i}SO(3)^{i}$. Consequently, the
natural stage for autonomous Lagrangian dynamics of robot motion
is the {tangent
bundle} $TQ_{rob}$, defined as follows.
To each $n-$dimensional ($n$D) {configuration
manifold} $Q$ there is associated its $2n$D {velocity phase--space
manifold}, denoted by $TQ$ and called the tangent bundle of $Q$.
The original smooth manifold $Q$ is called the {base} of $TQ$.
There is an onto map $\pi :TQ\rightarrow Q$, called the
{projection}. Above each point $x\in Q$ there is a {tangent space}
$T_{x}Q=\pi ^{-1}(x)$ to $Q$ at $x$, which
is called a {fibre}. The fibre $T_{x}Q\subset TQ$ is the subset of $%
TQ $, such that the total tangent bundle,
$TQ=\dbigsqcup\limits_{m\in Q}T_{x}Q$, is a {disjoint union} of
tangent spaces $T_{x}Q$ to $Q$ for all points $x\in Q$. From
dynamical perspective, the most important quantity in the tangent
bundle concept is the smooth map $v:Q\rightarrow TQ$, which
is an inverse to the projection $\pi $, i.e, $\pi \circ v=\func{Id}%
_{Q},\;\pi (v(x))=x$. It is called the {velocity vector--field}.
Its graph $(x,v(x))$ represents the {cross--section} of the
tangent
bundle $TQ$. This explains the dynamical term {velocity phase--space}%
, given to the tangent bundle $TQ$ of the manifold $Q$. The tangent bundle
is where tangent vectors live, and is itself a smooth manifold.
Vector--fields are cross-sections of the tangent bundle.
Robot's \emph{Lagrangian} (energy function) is a natural energy
function on the tangent bundle $TQ$.\footnote{
The corresponding autonomous
Hamiltonian robot dynamics takes place in the {cotangent bundle} $T^{\ast }Q_{rob}$, defined as follows.
A {dual} notion to the tangent space $T_{m}Q$ to a smooth manifold
$Q$ at a point $m$ is its {cotangent space} $T_{m}^{\ast }Q$ at
the same point $m$. Similarly to the tangent bundle, for a smooth
manifold $Q$ of dimension $n$, its {cotangent bundle} $T^{\ast }Q$
is the disjoint union of all its cotangent spaces $T_{m}^{\ast }Q$
at all points $m\in Q$, i.e., $T^{\ast }Q=\dbigsqcup\limits_{m\in
Q}T_{m}^{\ast }Q$. Therefore, the cotangent bundle of an
$n-$manifold $Q$ is the vector bundle $T^{\ast }Q=(TQ)^{\ast }$,
the (real) dual of the tangent bundle $TQ$. The cotangent bundle
is where 1--forms live, and is itself a smooth manifold.
Covector--fields (1--forms) are cross-sections of the cotangent
bundle. Robot's \emph{Hamiltonian} is a natural energy function on the
cotangent bundle.}

\subsection{Realistic Human Configuration Manifold}

\begin{figure}[tbh]
\centering \includegraphics[width=13cm]{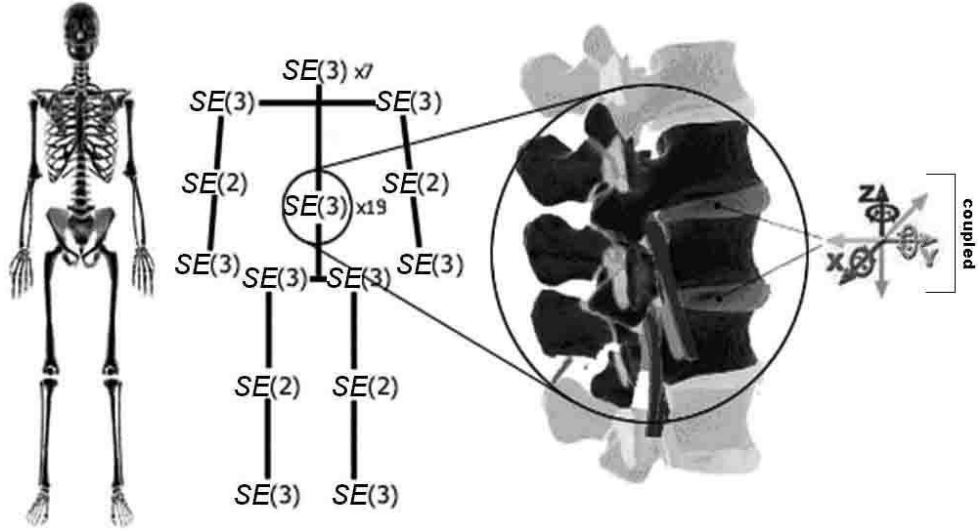} \caption{The
configuration manifold $Q$ of the human musculo-skeletal dynamics is defined as a
topological product of constrained $SE(3)$ groups acting in all
major (synovial) human joints, $Q=\prod_{i}SE(3)^{i}$. The manifold }
\label{SpineSE(3)}
\end{figure}

On the other hand, human joints are more flexible than robot joints. Namely,
every rotation in all synovial human joints is followed by the corresponding
micro--translation, which occurs after the rotational amplitude is reached
\cite{TijIJHR}. So, representation of human motion is rigorously defined
in terms of {Euclidean} $SE(3)$--groups of full rigid--body motion \cite%
{Marsden,GaneshSprSml,GaneshSprBig,GaneshADG} in all main human joints (see
Figure \ref{SpineSE(3)}). Therefore, the configuration manifold $Q$
for human dynamics is defined as a topological product of all included
constrained $SE(3)$ groups, $Q=\prod_{i}SE(3)^{i}$. Consequently, the
natural stage for autonomous Lagrangian dynamics of human motion is the
tangent bundle $TQ$ (and for the corresponding
autonomous Hamiltonian dynamics is the cotangent bundle $T^{\ast }Q$).

Briefly, the Euclidean SE(3)--group is defined as a semidirect
(noncommutative) product (denoted by $\rhd$) of 3D rotations and 3D translations: ~$%
SE(3):=SO(3)\rhd \mathbb{R}^{3}$. Its most important subgroups are the
following (for technical details see \cite%
{GaneshSprBig,ParkChung,GaneshADG}):\\

{{\frame{$%
\begin{array}{cc}
\mathbf{Subgroup} & \mathbf{Definition} \\ \hline
\begin{array}{c}
SO(3),\text{ group of rotations} \\
\text{in 3D (a spherical joint)}%
\end{array}
&
\begin{array}{c}
\text{Set of all proper orthogonal } \\
3\times 3-\text{rotational matrices}%
\end{array}
\\ \hline
\begin{array}{c}
SE(2),\text{ special Euclidean group} \\
\text{in 2D (all planar motions)}%
\end{array}
&
\begin{array}{c}
\text{Set of all }3\times 3-\text{matrices:} \\
\left[
\begin{array}{ccc}
\cos \theta & \sin \theta & r_{x} \\
-\sin \theta & \cos \theta & r_{y} \\
0 & 0 & 1%
\end{array}%
\right]%
\end{array}
\\ \hline
\begin{array}{c}
SO(2),\text{ group of rotations in 2D} \\
\text{subgroup of }SE(2)\text{--group} \\
\text{(a revolute joint)}%
\end{array}
&
\begin{array}{c}
\text{Set of all proper orthogonal } \\
2\times 2-\text{rotational matrices} \\
\text{ included in }SE(2)-\text{group}%
\end{array}
\\ \hline
\begin{array}{c}
\mathbb{R}^{3},\text{ group of translations in 3D} \\
\text{(all spatial displacements)}%
\end{array}
& \text{Euclidean 3D vector space}%
\end{array}%
$}}}\bigskip

The configuration manifold $Q=\prod_{i}SE(3)^{i}$ has the Riemannian geometry with the \textit{local metric form}: $$\langle g\rangle\equiv ds^{2}=g_{ij}dx^{i}dx^{j},\qquad\text{(Einstein's summation convention is in use)}$$
where $g_{ij}(x)$ is the material metric tensor defined by the biomechanical system's \emph{mass-inertia matrix} and $dx^{i}$
are differentials of the local joint coordinates $x^i$ on $Q$. Besides giving the local
distances between the points on the manifold
$Q,$ the Riemannian metric form $\langle g\rangle$
defines the system's kinetic energy: $$T=\frac{1}{2}g_{ij}\dot{x}^{i}\dot{x}^{j},$$
giving the \emph{Lagrangian equations} of the conservative skeleton motion with kinetic-minus-potential energy Lagrangian $L=T-V$, with the corresponding \emph{geodesic form} \cite{TijNL}
\begin{equation}
\frac{d}{dt}L_{\dot{x}^{i}}-L_{x^{i}}=0\qquad\text{or}\qquad \ddot{x}^i+\Gamma _{jk}^{i}\dot{x}^{j}\dot{x}^{k}=0, \label{geodes}
\end{equation}%
where subscripts denote partial derivatives, while $\Gamma _{jk}^{i}$ are the Christoffel symbols of
the affine Levi-Civita connection of the biomechanical manifold $Q$.

This is the basic geometrical structure for \emph{autonomous Lagrangian biomechanics}. In the next section will extend this basic structure to embrace the time-dependent biomechanics.
\begin{figure}[h]
 \centerline{\includegraphics[width=11cm]{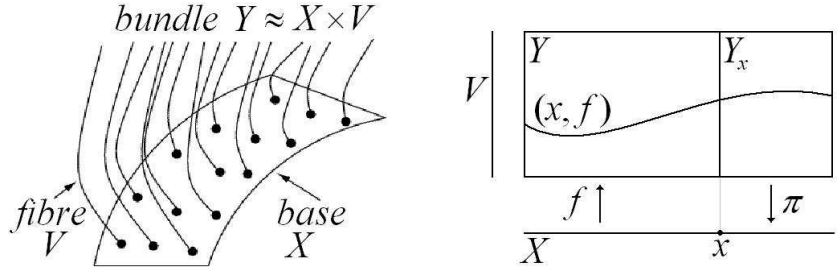}}
\caption{A sketch of a fibre bundle $Y\approx X\times V$ as a
generalization of a product space $X\times V$; left -- main
components; right -- a few details (see text for
explanation).}\label{Fibre1}
\end{figure}

\section{Biomechanical Jet Spaces}

In general, tangent and cotangent bundles, $TM$ and $T^{\ast }M$, of a smooth manifold $M$, are special cases of a more general geometrical object called
\emph{fibre bundle}, where the word \emph{fiber} $V$ of a map $\pi
:Y\rightarrow X$ denotes the \emph{preimage} $\pi^{-1}(x)$ of an
element $x\in X$. It is a space which \emph{locally} looks like a
product of two spaces (similarly as a manifold locally looks like
Euclidean space), but may possess a different \emph{global}
structure. To get a visual intuition behind this fundamental
geometrical concept, we can say that a fibre bundle $Y$ is a
\emph{homeomorphic generalization} of a \emph{product space}
$X\times V$ (see Figure \ref{Fibre1}), where $X$ and $V$ are
called the \emph{base} and the \emph{fibre}, respectively. $\pi
:Y\rightarrow X$ is called the \emph{projection}, $Y_{x}=\pi
^{-1}(x)$ denotes a fibre over a point $x$ of the base $X$, while
the map $f=\pi ^{-1}:X\rightarrow Y$ defines the
\emph{cross--section}, producing the \textit{graph} $(x,f(x))$ in
the bundle $Y$ (e.g., in case of a tangent bundle, $f=\dot{x}$
represents a velocity vector--field).

The main reason why we need to study fibre bundles is that
\emph{all dynamical objects} (including vectors, tensors,
differential forms and gauge potentials) are their
\emph{cross--sections}, representing \emph{generalizations of
graphs of continuous functions}.

\begin{figure}[h]
\centerline{\includegraphics[width=7cm]{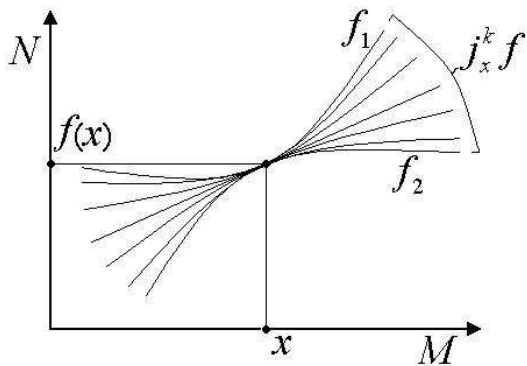}} \caption{An
intuitive geometrical picture behind the $k-$jet concept, based on
the idea of a higher--order tangency (or, higher--order contact). }
\label{jet1}
\end{figure}

By extending this line of formal bundle thinking, we come to the concept of a \textit{jet manifold}, which is based on the idea of \textit{higher--order tangency}, or higher--order
contact, at some designated point on a smooth manifold. Namely, a
pair of smooth manifold maps, ~$f_{1},f_{2}:M\rightarrow N$~ (see
Figure \ref{jet1}), are said to be $k-$\emph{tangent} (or
\emph{tangent of order }$k$, or
have a $k$th \emph{order contact}) at a point $x$ on a domain manifold $M$, denoted by $%
f_{1}\sim f_{2}$, iff
\begin{eqnarray*}
f_{1}(x) &=&f_{2}(x)\qquad \text{called}\qquad 0-\text{tangent}, \\
\partial _{x}f_{1}(x) &=&\partial _{x}f_{2}(x),\qquad \text{called}\qquad 1-%
\text{tangent}, \\
\partial _{xx}f_{1}(x) &=&\partial _{xx}f_{2}(x),\qquad \text{called}\qquad 2-%
\text{tangent}, \\
&&...\qquad \text{etc. to the order }k
\end{eqnarray*}

In this way defined $k-$\emph{tangency} is an \emph{equivalence
relation}, i.e.,
\[
f_{1}\sim f_{2}\Rightarrow f_{2}\sim f_{1},\qquad f_{1}\sim
f_{2}\sim f_{3}\Rightarrow f_{1}\sim f_{3},\qquad f_{1}\sim f_{1.}
\]

Now, a $k-$\textit{jet} (or, a \emph{jet of order }$k$), denoted
by $j_{x}^{k}f$, of a smooth map $f:M\rightarrow N$ at a
point $x\in M$ (see Figure \ref {jet1}), is defined as an
\emph{equivalence class} of $k-$tangent maps at $x$,
\begin{equation*}
j_{x}^{k}f:M\rightarrow N=\{f':f'\text{ is }k-\text{tangent to
}f\text{ at }x\}
\end{equation*}

In the same way as in the case of a map $f:M\rightarrow
N,~x\mapsto f(x)$, also in the case of the $k-$jet
~$j_{x}^{k}f: M\rightarrow N,~x\mapsto f(x)$, the point $x$
in the \textit{domain} $M$ is called the \textit{source} of
$j_{x}^{k}f$ and the point $f(x)$ in the \textit{codomain} $N$ is
the \textit{target} of $j_{x}^{k}f$.

We choose local coordinates on $M$ and $N$ in the neighborhood of
the points $x$ and $f(x)$, respectively. Then the $k-$jet
$j_{x}^{k}f$ of any map close to $f$, at any point close to $x$,
can be given by its Taylor--series expansion at $x$, with
coefficients up to degree $k$. Therefore, in a fixed coordinate
chart, the $k-$jet is given by: $$j_{x}^{k}f:M\rightarrow
N\equiv~\{~\text{collection of Taylor coefficients up to degree}~
k~\}.$$
The set of all $k-$jets from $M$ to $N$ is called the
\textit{$k-$jet space} $J^{k}(M,N)$. It has a natural
smooth--manifold structure. Also, a map from a \textit{$k-$jet
manifold} $J^{k}(M,N)$ to a smooth manifold $M$ or $N$ is called a
\textit{$k-$jet bundle}.

For example, consider a simple function
~$f:X\rightarrow Y,\,x\mapsto y=f(x)$, mapping the $X-$axis
into the $Y-$axis in $\mathbb{R}^2$. In this case, $X$ is a domain
and $Y$ is a codomain. At a chosen point
$x\in X$ we have:\\ a $0-$jet is a graph: $(x,f(x))$;\\
a $1-$jet is a triple: $(x,f(x),f{'}(x))$;\\ a
$2-$jet is a quadruple: $(x,f(x),f{'}(x),f^{\prime \prime }(x))$,\\
~~ and so on, up to the order $k$ (where
$f{'}(x)=\frac{df(x)}{dx}$, etc).\\ The set of all $k-$jets from
$j^k_xf:X\rightarrow Y$ is called the $k-$jet manifold
$J^{k}(X,Y)$.

We now turn back into the field of time-dependent human biomechanics, where the fundamental geometrical construct is the \emph{configuration fibre bundle}
$\pi:Q\rightarrow \mathbb{R}$.
Given a configuration fibre bundle $Q\rightarrow \mathbb{R}$ over
the time axis $\mathbb{R}$, we say that the \textit{$1-$jet
manifold} $J^{1}(\mathbb{R},Q)$
is the set of equivalence classes $j_{t}^{1}s$ of sections $s^{i}:\mathbb{R}%
\rightarrow Q$ of the bundle $Q\rightarrow \mathbb{R}$, which are
identified by their values $s^{i}(t)$, as well as by the values of their partial derivatives $%
\partial _{t}s^{i}=\partial _{t}s^{i}(t)$ at time points $t\in \mathbb{R}$.
The 1--jet manifold $J^{1}(\mathbb{R},Q)$ is coordinated by $(t,x^{i},\dot{x}%
^{i})$, so the 1--jets are local coordinate maps
\begin{equation*}j_{t}^{1}s:\mathbb{R}%
\rightarrow Q,\qquad t\mapsto (t,x^{i},\dot{x}^{i})
\end{equation*}
Similarly, the \textit{$2-$jet manifold} $J^{2}(\mathbb{R},Q)$
is the set of equivalence classes $j_{t}^{2}s$ of sections $s^{i}:\mathbb{R}\rightarrow Q$%
\ of the configuration bundle $\pi:Q\rightarrow \mathbb{R}$, which
are identified by their values $s^{i}(t)$, as well as the values
of their first and second partial derivatives, $\partial
_{t}s^{i}=\partial _{t}s^{i}(t)$
and $\partial _{tt}s^{i}=\partial _{tt}s^{i}(t)$, respectively, at time points $%
t\in \mathbb{R}$. The 2--jet manifold $J^{2}(\mathbb{R},Q)$ is
coordinated by $(t,x^{i},\dot{x}^{i},\ddot{x}^{i})$, so the
2--jets are local coordinate maps
\begin{equation*}j_{t}^{2}s:\mathbb{R}%
\rightarrow Q,\qquad t\mapsto
(t,x^{i},\dot{x}^{i},\ddot{x}^{i}).\end{equation*}

\begin{figure}[h]
\centerline{\includegraphics[width=4cm]{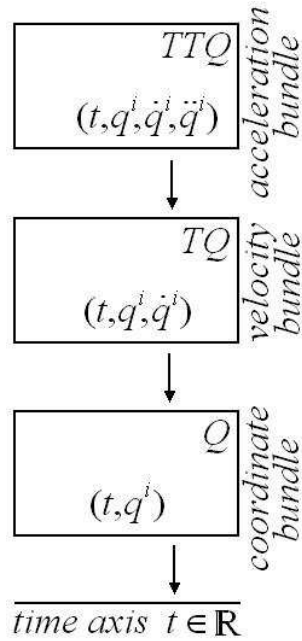}}
\caption{Hierarchical geometrical structure of time--dependent
biomechanics. Here, generalized coordinates $q^i$, velocities $\dot{q}^i$ and accelerations $\ddot{q}^i$ replace the corresponding rotational and translational joint coordinates $x^i$, velocities $\dot{x}^i$ and accelerations $\ddot{x}^i$.} \label{GeoMechStr}
\end{figure}

This global geometrical structure of time--dependent biomechanics is
depicted in Figure \ref{GeoMechStr}.

In the framework of biomechanics, we consider a pair of maps
$f_{1},f_{2}:\mathbb{R}\rightarrow Q $ from the real line
$\mathbb{R}$, representing the {time} $t-$axis, into a smooth $n-$dimensional Riemannian
{configuration manifold} $Q$ from Figure \ref{SpineSE(3)}. We say that the two
maps $f_{1}=f_{1}(t)$ and $f_{2}=f_{2}(t)$ have the same $k-$jet {$%
j_{t}^{k}f $} at a specified time instant $t_{0}\in \mathbb{R}$, iff:
\begin{enumerate}
\item $f_{1}(t)=f_{2}(t)$ at $t_{0}\in \mathbb{R}$; and also
\item the first $k$ terms of their Taylor--series expansions around $%
t_{0}\in \mathbb{R}$ are equal.
\end{enumerate}
The set of all $k-$jets $j_{t}^{k}f:\mathbb{R}\rightarrow Q$ is the $k-$jet
manifold $J^{k}(\mathbb{R},Q)$. In particular, $J^{1}(\mathbb{R},Q)\cong
\mathbb{R}\times TQ$.

\subsection{Jet Prolongation of Locomotion Vector-fields}

Consider an arbitrary \emph{locomotion vector-field} $u$ that is on the \emph{biomechanical
configuration bundle} $\pi:Q\rightarrow \mathbb{R}$ defined by
\begin{equation*}
u=u^{t}\partial _{t}+u^{i}(t,q^{j})\partial _{i}.
\end{equation*}
The so-called \emph{jet prolongation} of the locomotion vector-field $u$ onto the \emph{extended velocity phase-space}, tht is the 1--jet manifold $J^{1}(\mathbb{R},Q)$, reads
\begin{equation*}
J^{1}u=u^{t}\partial _{t}+u^{i}\partial _{i}+d_{t}u^{i}\partial _{i}^{t}.
\end{equation*}

For example, consider a one-dimensional motion of a point particle subject
to friction. It is described by the dynamic equation \
\begin{equation*}
\ddot{q}=-k\dot{q},\qquad (\text{ with \ }k>0).
\end{equation*}
This is the Lagrangian equation for the Lagrangian function
\begin{equation*}
L=\frac{1}{2}\exp [kt]\dot{q}^{2}dt.
\end{equation*}
Let us consider the vector field
\begin{equation*}
v=\partial _{t}-\frac{k}{2}q\partial _{q}.
\end{equation*}
Its jet prolongation $J^{1}v$ reads
\begin{equation*}
J^{1}v=\partial _{t}-\frac{k}{2}q\partial _{q}-\frac{k}{2}q_{t}\partial
_{q}^{t}.
\end{equation*}\\

Generalization of all above jet structures to the $k-$\emph{jet manifold} $J^{k}(\mathbb{R%
},Q)$ is obvious,
\begin{equation*}j_{t}^{k}s:\mathbb{R}%
\rightarrow Q,\qquad t\mapsto
(t,x^{i},\dot{x}^{i},\ddot{x}^{i},\dddot{x}^{i},...,x^{(k)i}).
\end{equation*}
For more technical details on jet manifolds and bundles with their applications to mechanics and physics, see \cite{Saunders,massa,book,sard98,book99}).

\section{Lagrangian Form of Time-Dependent Biomechanics}

The general form of time-dependent Lagrangian biomechanics with \emph{time-dependent
Lagrangian} function $L(t;x^{i};\dot{x}^{i})$ defined on the 1--jet
manifold $X=J^{1}(\mathbb{R},Q)\cong
\mathbb{R}\times TQ$, with local canonical coordinates:
$(t;x^{i};\dot{x}^{i})=$ (time, coordinates and velocities) in active human joints, can be formulated as \cite{GaneshSprBig,GaneshADG}
\begin{equation}
\frac{d}{dt}L_{\dot{x}^{i}}-L_{x^{i}}=\mathcal{F}_{i}\left( t,x,\dot{x}%
\right) ,\qquad (i=1,...,n),  \label{classic}
\end{equation}%
where the coordinate and velocity partial derivatives of the Lagrangian are
respectively denoted by $L_{x^{i}}$ and $L_{\dot{x}^{i}}$.
The right--hand side terms $\mathcal{F}_{i}(t,x,\dot{x})$ of
(\ref{classic}) denote any type of {external} torques and forces,
including \cite{GaneshSprSml,GaneshWSc}: (i) excitation and contraction dynamics of
muscular--actuators; (ii) autogenetic (spinal) reflex force controllers; (iii) rotational dynamics of hybrid robot
actuators; (iv) (nonlinear) dissipative joint torques and
forces; and (v) external stochastic perturbation torques and forces.

\subsection{Time-Dependent Riemannian Geometry and Ricci Flow}

In the geodesic framework (\ref{geodes}), the (in)stability of the
biomechanical joint and center-of-mass trajectories is the (in)stability of the geodesics, and it is
completely determined by the curvature properties of the
underlying manifold according to the \textit{Jacobi equation} of
\emph{geodesic deviation} \cite{GaneshSprBig,GaneshADG}
\begin{equation*}
\frac{D^{2}J^{i}}{ds^{2}}+R_{~jkm}^{i}\frac{dx^{j}}{ds}J^{k}\frac{dx^{m}}{ds}%
=0,
\end{equation*}%
whose solution $J$, usually called \textit{Jacobi variation
field}, locally measures the distance between nearby geodesics;
$D/ds$ stands for the \textit{covariant derivative} along a
geodesic and $R_{~jkm}^{i}$ are the components of the
\textit{Riemann curvature tensor}.

On the other hand, the mass-inertia matrix of human body segments, defining the Riemannian metric tensor $g_{ij}=g_{ij}(x)$, need not be time-constant, that is, in general we have $g_{ij}=g_{ij}(x)=g_{ij}(t,x)$. Majority of fast movements in gymnastics are based on fast-changing mass distribution. This time-dependent Riemannian geometry can be formalized in terms of the
\textit{Ricci flow equation} (or, the parabolic
Einstein equation), introduced by R. Hamilton in 1982 \cite{Ham82}, that is the
nonlinear heat--like evolution equation\footnote{%
The current hot topic in geometric topology is the Ricci flow, a Riemannian
evolution machinery that recently allowed G. Perelman to prove the
celebrated \textit{Poincar\'{e} Conjecture}, a century--old mathematics
problem and win him the 2006 Fields Medal (which he
declined in a public controversy). The Poincar\'{e}
Conjecture can roughly be put as a question: Is a closed 3--manifold $Q$
topologically a sphere if every closed curve in $Q$ can be shrunk
continuously to a point? In other words, Poincar\'{e} conjectured: A
simply-connected compact 3--manifold is diffeomorphic to the 3--sphere $%
S^{3} $.}
\begin{equation}
\partial _{t}g_{ij}=-2R_{ij},  \label{RF}
\end{equation}%
for a time--dependent Riemannian metric $g=g_{ij}(t)$ on a smooth $n-$manifold $Q$ with the Ricci curvature tensor $%
R_{ij} $. This equation roughly says
that we can deform any metric on a 2--surface or $n-$manifold by the
negative of its curvature; after \emph{normalization}, the final state of such deformation will be a metric with constant
curvature. The factor of 2 in (\ref{RF}) is more-or-less arbitrary, but the
negative sign is essential to insure a kind of global \emph{volume
exponential decay},\footnote{%
This complex geometric process is globally similar to a generic exponential
decay ODE:
\begin{equation*}
\dot{x}=-\lambda f(x),
\end{equation*}%
for a positive function $f(x)$. We can get some insight into its solution
from the simple exponential decay ODE,
\begin{equation*}
\dot{x}=-\lambda x\qquad \text{with the solution}\qquad x(t)=x_{0}\mathrm{e}%
^{-\lambda t},
\end{equation*}%
(where $x=x(t)$ is the observed quantity with its initial value $x_{0}$ and $%
\lambda $ is a positive decay constant), as well as the corresponding $n$th
order rate equation (where $n>1$ is an integer),%
\begin{equation*}
\dot{x}=-\lambda x^{n}\qquad \text{with the solution}\qquad \frac{1}{x^{n-1}}%
=\frac{1}{{x_{0}}^{n-1}}+(n-1)\,\lambda t.
\end{equation*}%
} since the Ricci flow equation (\ref{RF}) is a kind of nonlinear
geometric generalization of the standard linear \emph{heat
equation}
\begin{equation}
\partial _{t}u=\Delta u.  \label{h1}
\end{equation}%
Like the heat equation (\ref{h1}), the Ricci flow equation (\ref{RF}) is
well behaved in forward time and acts as a kind of smoothing operator (but
is usually impossible to solve in backward time). If some parts of a solid
object are hot and others are cold, then, under the heat equation, heat will
flow from hot to cold, so that the object gradually attains a uniform
temperature. To some extent the Ricci flow behaves similarly, so that the
Ricci curvature `tries' to become more uniform \cite{Milnor}, thus
resembling a monotonic \emph{entropy growth},\footnote{%
Note that two different kinds of entropy functional have been introduced
into the theory of the Ricci flow, both motivated by concepts of entropy in
thermodynamics, statistical mechanics and information theory. One is
Hamilton's entropy, the other is Perelman's entropy. While in Hamilton's
entropy, the scalar curvature $R$ of the metric $g_{ij}$ is viewed as the
leading quantity of the system and plays the role of a probability density,
in Perelman's entropy the leading quantity describing the system is the
metric $g_{ij}$ itself. Hamilton established the monotonicity of his entropy
along the volume-normalized Ricci flow on the 2--sphere $S^{2}$ \cite%
{surface}. Perelman established the monotonicity of his entropy along the
Ricci flow in all dimensions \cite{Perel1}.} $\partial _{t}S\geq 0$, which
is due to the positive definiteness of the metric $g_{ij}\geq 0$, and
naturally implying the \emph{arrow of time} \cite{GaneshADG}.

In a suitable local coordinate system, the Ricci flow equation (\ref{RF}) on a biomechanical configuration manifold $Q$
has a nonlinear heat--type form, as follows. At any time $t$, we can choose
local harmonic coordinates so that the coordinate functions are locally
defined harmonic functions in the metric $g(t)$. Then the Ricci flow takes
the general form \cite{Anderson}
\begin{eqnarray}
\partial _{t}g_{ij}&=&\Delta _{Q}g_{ij}+G_{ij}(g,\partial g), \qquad\text{where}  \label{RH} \\
\Delta _{Q}&\equiv & \frac{1}{\sqrt{\det (g)}}\frac{\partial }{\partial x^{i}}%
\left( \sqrt{\det (g)}g^{ij}\frac{\partial }{\partial
x^{j}}\right) \notag
\end{eqnarray} is the \textit{Laplace--Beltrami operator} of the configuration manifold $Q$
and $G_{ij}(g,\partial g)$ is a lower--order term
quadratic in $g$ and its first order partial derivatives $\partial
g$. From the analysis of nonlinear heat PDEs, one obtains
existence and uniqueness of forward--time solutions to the Ricci
flow on some time interval, starting at any smooth initial metric
$g_{0}$ on $Q$.

The exponentially-decaying geometrical diffusion (\ref{RH}) is a formal description for pirouettes in ice skating and fast rotational movements in gymnastics.

\subsection{Topological Lagrangian Dynamics}

The \emph{biomechanical jet
space} $X=J^{1}(\mathbb{R},X)\cong
\mathbb{R}\times TQ$ gives rise to the fundamental $n-$\emph{groupoid}, or $n-$category $\Pi _{n}(X)$ (see
\cite{GaneshSprBig,GaneshADG}). In $\Pi _{n}(X)$, 0--cells are
\emph{points} in $X$; 1--cells are \emph{paths} in
$X$ (i.e.,
parameterized smooth maps $f:[0,1]\rightarrow X$); 2--cells are \emph{%
smooth homotopies} (denoted by $\simeq $) \emph{of paths} relative
to endpoints (i.e., parameterized smooth maps $h:[0,1]\times
\lbrack 0,1]\rightarrow X$); 3--cells are \emph{smooth
homotopies of homotopies} of paths in $X$ (i.e.,
parameterized smooth maps $j:[0,1]\times \lbrack 0,1]\times
\lbrack 0,1]\rightarrow X$). Categorical \emph{composition}
is defined by \emph{pasting} paths and homotopies. In this way,
the following \textit{recursive Lagrangian homotopy dynamics} emerges on the
configuration manifold $X$:
\begin{eqnarray*}
&&\mathtt{0-cell:}\,\,x_{0}\,\node\,\,\,\qquad x_{0}\in X; \qquad
\text{in
the higher cells below: }t,s\in[0,1]; \\
&&\mathtt{1-cell:}\,\,x_{0}\,\node\cone{L}\node\,x_{1}\qquad
L:x_{0}\simeq
x_{1}\in X, \\
&&L:[0,1]\rightarrow X,\,L:x_{0}\mapsto
x_{1},\,x_{1}=L(x_{0}),\,L(0)=x_{0},\,L(1)=x_{1}; \\
&&\text{e.g., ~~linear path: }~~L(t)=(1-t)\,x_{0}+t\,x_{1};\qquad \text{or} \\
&&\text{Lagrangian }L-\text{dynamics with endpoint conditions }%
(x_0,x_1): \\
&&\frac{d}{dt}L_{\dot{x}^{i}}=L_{x^{i}},\qquad \text{with}\qquad
x(0)=x_{0},\qquad x(1)=x_{1},\qquad (i=1,...,n); \\
&&\mathtt{2-cell:}\,\,x_{0}\,\node\ctwodbl{L^1}{L^2}{h}\node\,x_{1}\qquad
h:L^1\simeq L^2\in X, \\
&&h:[0,1]\times \lbrack 0,1]\rightarrow X,\,h:L^1\mapsto L^2,\,L^2=h(L^1(x_{0})), \\
&&h(x_{0},0)=L^1(x_{0}),\,h(x_{0},1)=L^2(x_{0}),\,h(0,t)=x_{0},\,h(1,t)=x_{1} \\
&&\text{e.g., linear homotopy: }h(x_{0},t)=(1-t)\,L(x_{0})+t\,L^2(x_{0});\qquad%
\text{or} \\
&&\text{homotopy between two Lagrangian
}(L^1,L^2)-\text{dynamics}
\\
&&\text{with the same endpoint conditions }(x_0,x_1): \\
&&\frac{d}{dt}L^1_{\dot{x}^{i}}=L^1_{x^{i}},\qquad \text{and} \qquad \frac{d}{dt}%
L^2_{\dot{x}^{i}}=L^2_{x^{i}}\qquad\text{with}\qquad x(0)=x_{0},\qquad
x(1)=x_{1};\\ &&- \text{ etc.}
\end{eqnarray*}

\section{Conclusion}

In this paper we have presented time-dependent generalization of an `ordinary' autonomous human musculo-skeletal biomechanics. Firstly, we have defined the basic configuration manifold $Q$ of human musculo-skeletal biomechanics as an anthropomorphic chain of constrained Euclidean motion groups $SE(3)$. Secondly, we have extended this base manifold by the real time axis $\mathbb{R}$ into the biomechanical configuration bundle $Q\to\mathbb{R}$. The time-dependent Lagrangian biomechanics is defined on $Q\to\mathbb{R}$ using the formalism of first and second order jet manifolds as well as jet prolongations.
Then we moved to time-dependent Riemannian geometry, governed by the Ricci-flow diffusion and showed that the exponential-like decay of total biomechanical energy (due to exhaustion of biochemical resources) is closely related to the Ricci-flow based geometrical diffusion.


\begin{thebibliography}{99}
\bibitem{GaneshADG} Ivancevic, V., Ivancevic, T., Applied Differential
Geometry: A Modern Introduction. World Scientific, Singapore, (2007)

\bibitem{GaneshSprSml} Ivancevic, V., Ivancevic, T., Human--Like
Biomechanics: A Unified Mathematical Approach to Human Biomechanics and
Humanoid Robotics. Springer, Dordrecht, (2006)

\bibitem{GaneshWSc} Ivancevic, V., Ivancevic, T., Natural Biodynamics. World Scientific, Singapore (2006)

\bibitem{GaneshSprBig} Ivancevic, V., Ivancevic, T., Geometrical Dynamics of Complex Systems: A Unified Modelling Approach to Physics, Control,
Biomechanics, Neurodynamics and Psycho-Socio-Economical Dynamics. Springer, Dordrecht, (2006)

\bibitem{StrAttr} Ivancevic, V., Ivancevic, T., High--Dimensional Chaotic
and Attractor Systems. Springer, Berlin, (2006)

\bibitem{TijIJHR} Ivancevic, V., Ivancevic, T., Human versus humanoid robot biodynamcis. Int. J. Hum. Rob. \textbf{5}(4), 699–-713, (2008)

\bibitem{TijNis} Ivancevic, T., Jovanovic, B., Djukic, M., Markovic, S., Djukic, N., Biomechanical Analysis of Shots and Ball Motion in Tennis and the Analogy with Handball Throws, J. Facta Universitatis, Series: Sport, \textbf{6}(1), 51--66, (2008)

\bibitem{TijNL} Ivancevic, T., Jain, L., Pattison, J., Hariz, A., Nonlinear Dynamics and Chaos Methods in Neurodynamics and Complex Data Analysis, Nonl. Dyn. (Springer), \textbf{56}(1-2), 23--44, (2009)

\bibitem{TijSpr} Ivancevic, T., Jovanovic, B., Djukic, S., Djukic, M., Markovic, S., Complex Sports Biodynamics: With Practical Applications in Tennis, Springer, Berlin, (2009)

\bibitem{SciSport} Tucker, R., Dugas, J.: Beijing 2008: Men 100\,m race analysis. Bolt's 9.69s. Analysis of speed during the world record. The Science of Sport, http://www.sportsscientists.com/2008/08/beijing-2008-men-100m-race-analysis.html, (2008)

\bibitem{Marsden} Marsden, J.E., Ratiu, T.S., Introduction to Mechanics and
Symmetry: A Basic Exposition of Classical Mechanical Systems. (2nd ed),
Springer, New York, (1999)

\bibitem{ParkChung} Park, J., Chung, W.-K., Geometric Integration on
Euclidean Group With Application to Articulated Multibody Systems. IEEE
Trans. Rob. \textbf{21}(5), 850--863 (2005)

\bibitem{Hatze} Hatze, H., A general myocybernetic control model of skeletal
muscle. Biol. Cyber. \textbf{28}, 143--157, (1978)

\bibitem{Wilkie} Wilkie, D.R., The mechanical properties of muscle. Brit.
Med. Bull. \textbf{12}, 177--182, (1956)

\bibitem{Hill} Hill, A.V.,The heat of shortening and the dynamic constants
of muscle. Proc. Roy. Soc. \textbf{B76}, 136--195, (1938)

\bibitem{Vuk} Vukobratovic, M., Borovac, B., Surla, D., Stokic, D., Biped
Locomotion: Dynamics, Stability, Control, and Applications. Springer,
Berlin, (1990)

\bibitem{NeuFuz} Ivancevic, V., Ivancevic, T., Neuro--Fuzzy Associative
Machinery for Comprehensive Brain and Cognition Modelling. Springer, Berlin, (2007)

\bibitem{Saunders} Saunders, D.J.: The Geometry of Jet Bundles. Lond. Math. Soc. Lect. Notes Ser. \textbf{142}, Cambr. Univ. Pr., (1989)

\bibitem{massa} Massa, E., Pagani, E., Jet bundle geometry, dynamical connections and the inverse problem of Lagrangian mechanics. Ann. Inst. Henri Poincar\'e {\bf 61}, 17, (1994)

\bibitem{book} Giachetta, G., Mangiarotti, L., Sardanashvily, G., New Lagrangian and Hamiltonian Methods in Field Theory, World Scientific, Singapore, (1997)

\bibitem{sard98} Sardanashvily, G.: Hamiltonian time-dependent mechanics. J. Math. Phys. {\bf 39}, 2714, (1998)

\bibitem{book99} Mangiarotti, L., Obukhov, Yu., Sardanashvily, G., Connections in Classical and Quantum Field Theory. World Scientific, Singapore, (1999)

\bibitem{sard93} Sardanashvily, G., Gauge Theory in Jet Manifolds. Hadronic Press, Palm Harbor, FL, (1993)

\bibitem{Ham82} {R.S. Hamilton}, Three-manifolds with positive Ricci
curvature, J. Diff. Geom. \textbf{17}, 255-306, (1982)

\bibitem{4-manifold} {R.S. Hamilton}, Four-manifolds with positive
curvature operator, J. Dif. Geom. \textbf{24}), 153-179, (1986)

\bibitem{surface} {R.S. Hamilton}, The Ricci flow on surfaces, Cont.
Math. \textbf{71}, 237-261, (1988)

\bibitem{Harnack} {R.S. Hamilton}, The Harnack estimate for the Ricci
flow, J. Dif. Geom. \textbf{37}, 225-243, (1993)

\bibitem{non-singular} {R.S. Hamilton}, Non-singular solutions of the
Ricci flow on three-manifolds, Comm. Anal. Geom. \textbf{7}(4), 695-729, (1999)

\bibitem{Milnor} {J. Milnor}, Towards the Poincar\'{e} Conjecture and
the Classification of 3-Manifolds, Not. Am. Math. Soc. \textbf{50}(10), 1226-1233, (2003)

\bibitem{Perel1} {G. Perelman}, The entropy formula for the Ricci
flow and its geometric applications, arXiv:math.DG/0211159, (2002)

\bibitem{Anderson} {M.T. Anderson}, Geometrization of 3-manifolds via
the Ricci flow, Not. Am. Math. Soc. \textbf{51}(2), 184-193, (2004)
\end{thebibliography}
\end{document}